\documentclass[sigconf]{acmart}
\usepackage{amsthm}
\usepackage{algpseudocode}% change to algorithm algorithm2e algorithmic
\usepackage{bm}
\usepackage{balance}
\usepackage{color}
\usepackage{colortbl}
\usepackage{float}
\usepackage{graphicx}
\usepackage{listings}
\usepackage{multirow}
\usepackage{url}
\usepackage{stfloats}
\usepackage{subcaption}

\makeatletter

\floatstyle{ruled}
\newfloat{algorithm}{tbp}{loa}
\providecommand{\algorithmname}{Algorithm}
\floatname{algorithm}{\protect\algorithmname}
\usepackage{tikz}
\usetikzlibrary{calc}

\theoremstyle{plain}

\theoremstyle{plain}

\theoremstyle{plain}

\theoremstyle{plain}

\usepackage[acronym]{glossaries}
\newcommand{\newac}{\newacronym}
\newcommand{\ac}{\gls}
\newcommand{\Ac}{\Gls}
\newcommand{\acpl}{\glspl}

\newac{speb}{SPEB}{square position error bound}
\newac[plural=EFIMs,firstplural=Fisher information matrices (EFIMs)]{efim}{EFIM}{Fisher information matrix}
\newac{ne}{NE}{Nash equilibrium}
\newac{mse}{MSE}{mean squared error}
\newac{toa}{TOA}{time-of-arrival}
\newac{snr}{SNR}{signal-to-noise ratio}
\newac{lan}{LAN}{local area network}
\newac{psd}{PSD}{positive semidefinite}
\newac{pd}{PD}{positive definite}
\newac{wrt}{w.r.t.}{with respect to}
\newac{lhs}{L.H.S.}{left hand side}
\newac{wp1}{w.p.1}{with probability 1}
\newac{kkt}{KKT}{Karush-Kuhn-Tucker}
\newac{wlog}{w.l.o.g.}{without loss of generality}
\newac{mle}{MLE}{maximum likelihood estimation}
\newac{gps}{GPS}{Global Positioning System}
\newac{rssi}{RSSI}{received signal strength indicator}
\newac{mimo}{MIMO}{multiple-input multiple-output}
\newac{csi}{CSI}{channel state information}
\newac{fdd}{FDD}{frequency division duplexing}
\newac{ms}{MS}{mobile station}
\newac{bs}{BS}{base station}
\newac{d2d}{D2D}{device-to-device}
\newac{slnr}{SLNR}{signal-to-interference-leakage-and-noise-ratio}
\newac{ula}{ULA}{uniform linear antenna array}
\newac{pas}{PAS}{power angular spectrum}
\newac{mmse}{MMSE}{minimum mean square error}
\newac{zf}{ZF}{zero-forcing}
\newac{rzf}{RZF}{regularized zero-forcing}
\newac{as}{AS}{angular spread}
\newac{aod}{AOD}{angle of departure}
\newac{iid}{i.i.d.}{independent and identically distributed} 
\newac{sinr}{SINR}{signal-to-interference-and-noise ratio}
\newac{tdd}{TDD}{time-division duplex}
\newac{rvq}{RVQ}{random vector quantization}
\newac{rhs}{R.H.S.}{right hand side}
\newac{mrc}{MRC}{maximum ratio combining}
\newac{cdf}{CDF}{cumulative distribution function}
\newac{a.s.}{a.s.}{almost surely}
\newac{los}{LOS}{line-of-sight}
\newac{jsdm}{JSDM}{joint spatial division and multiplexing}
\newac{map}{MAP}{maximum a posteriori}
\newac{klt}{KLT}{Karhunen-Lo\`eve Transform}
\newac{lbe}{LBE}{link bargaining equilibrium}
\newac{se}{SE}{Stackelberg equilibrium}
\newac{uav}{UAV}{unmanned aerial vehicle}
\newac{nlos}{NLOS}{non-line-of-sight}
\newac{pdf}{PDF}{probability density function}
\newac{em}{EM}{expectation-maximization}
\newac{knn}{KNN}{$k$-nearest neighbor}
\newac{svd}{SVD}{singular value decomposition}
\newac{nmf}{NMF}{non-negative matrix factorization}
\newac{umf}{UMF}{unimodality-constrained matrix factorization}
\newac{rmse}{RMSE}{rooted mean squared error}
\newac{olos}{OLOS}{obstructed line-of-sight}
\newac{mmw}{mmW}{millimeter wave}
\newac{ber}{BER}{bit error rate}
\newac{rss}{RSS}{received signal strength}
\newac{lp}{LP}{linear program}
\newac{ufw}{U-FW}{unimodal Frank-Wolfe}
\newac{utf}{UTF}{unimodality-constrained tensor factorization}
\newac{fw}{FW}{Frank-Wolfe}
\newac{iot}{IoT}{Internet-of-Things}
\newac{mae}{MAE}{mean absolute error}
\newac{crb}{CRB}{Cram\'er-Rao bound}
\newac{aoa}{AoA}{angle of arrival}
\newac{wcl}{WCL}{weighted centroid localization}
\newac[plural=GNSS,firstplural=global navigation satellite systems (GNSS)]{gnss}{GNSS}{global navigation satellite system}
\newac{gsm}{GSM}{Global System for Mobile Communications}
\newac{imu}{IMU}{inertial measurement unit}
\newac{rbf}{RBF}{radial basis function}
\newac{msf}{MSF}{multi-sensor fusion}
\newac{lidar}{LiDAR}{light detection and ranging}
\newac{glrt}{GLRT}{generalized likelihood ratio test}
\newac{sdr}{SDR}{software-defined radio}
\newac{ap}{AP}{access point}
\newac{lte}{LTE}{Long-Term Evolution}
\newac{raim}{RAIM}{receiver autonomous integrity monitoring}
\newac{pvt}{PVT}{position, velocity and time}
\newac{npl}{NPL}{Neyman-Pearson lemma}
\newac{ekf}{EKF}{extended Kalman filter}
\newac{tdoa}{TDOA}{time difference of arrival}
\newac{dop}{DOP}{dilution of precision}
\newac[plural=SOPs,firstplural=signals of opportunity (SOPs)]{sop}{SOP}{signals of opportunity}
\newac[plural=LBS,firstplural=location-based services (LBS)]{lbs}{LBS}{location-based service}
\newac{llm}{LLM}{large language model}
\newac{ssid}{SSID}{service set identifier}
\newac{bssid}{BSSID}{basic service set identifier}
\newac{mac}{MAC}{medium access control}
\newac[plural=POIs,firstplural=points of interest (POIs)]{poi}{POI}{point of interest}
\newac{uwb}{UWB}{ultra-wideband}
\newac{sbas}{SBAS}{satellite-based augmentation system}
\newac{rtt}{RTT}{round-trip time}
\newac{icmp}{ICMP}{Internet Control Message Protocol}
\newac{vpn}{VPN}{virtual private network}
\newac{roc}{ROC}{receiver-operating characteristic curve}

\makeatother

\providecommand{\corollaryname}{Corollary}
\providecommand{\lemmaname}{Lemma}
\providecommand{\propositionname}{Proposition}
\providecommand{\theoremname}{Theorem}

\definecolor{mycolor1}{rgb}{0.494117647058824,0.184313725490196,0.556862745098039}
\definecolor{mycolor2}{rgb}{0.466666666666667,0.674509803921569,0.188235294117647}
\definecolor{mycolor3}{rgb}{0.301960784313725,0.745098039215686,0.933333333333333}
\definecolor{mycolor4}{rgb}{0.929411764705882,0.694117647058824,0.125490196078431}
\definecolor{mycolor5}{rgb}{0.635294117647059,0.078431372549020,0.184313725490196}
\definecolor{mycolor6}{rgb}{0.8500,0.3250,0.0980}
\copyrightyear{2025}
\acmYear{2025}
\setcopyright{rightsretained}
\acmConference[WiSec 2025] {Proceedings of the 18th ACM Conference on Security and Privacy in Wireless and Mobile Networks}{ June 30--July 3, 2025}{Arlington, VA, USA.}
\acmBooktitle{Proceedings of the 18th ACM Conference on Security and Privacy in Wireless and Mobile Networks (WiSec 2025), June 30--July 3, 2025, Arlington, VA, USA}
\acmISBN{979-8-4007-1530-3/25/06}
\acmDOI{10.1145/3734477.3734707}
\title{Guardian Positioning System (GPS) for Location Based Services}% Using Opportunistic Information

\author{Wenjie Liu}
\affiliation{%
  \institution{Networked Systems Security (NSS) Group}
  \institution{KTH Royal Institute of Technology}
  \city{Stockholm}
  \country{Sweden}}
\email{wenjieli@kth.se}

\author{Panos Papadimitratos}
\affiliation{%
  \institution{Networked Systems Security (NSS) Group}
  \institution{KTH Royal Institute of Technology}
  \city{Stockholm}
  \country{Sweden}}
\email{papadim@kth.se}

\begin{document}
\begin{abstract}
\Ac{lbs} applications proliferate and support transportation, entertainment, and more. Modern mobile platforms, with smartphones being a prominent example, rely on terrestrial and satellite infrastructures (e.g., \ac{gnss} and crowdsourced Wi-Fi, Bluetooth, cellular, and IP databases) for correct positioning. However, they are vulnerable to attacks that manipulate positions to control and undermine \ac{lbs} functionality---thus enabling the scamming of users or services. Our work reveals that \ac{gnss} spoofing attacks succeed even though smartphones have multiple sources of positioning information. Moreover, that Wi-Fi spoofing attacks with \ac{gnss} jamming are surprisingly effective. More concerning is the evidence that sophisticated, coordinated spoofing attacks are highly effective. Attacks can target \ac{gnss} in combination with other positioning methods, thus defenses that assume that only \ac{gnss} is under attack cannot be effective. More so, resilient \ac{gnss} receivers and special-purpose antennas are not feasible on smartphones. To address this gap, we propose an extended \ac{raim} framework that leverages the readily available, redundant, often so-called opportunistic positioning information on off-the-shelf platforms. We jointly use onboard sensors, terrestrial infrastructures, and \ac{gnss}. We show that our extended \ac{raim} framework improves resilience against location spoofing, e.g., achieving a detection accuracy improvement of up to 24--58\% compared to the state-of-the-art algorithms and location providers; detecting attacks within 5 seconds, with a low false positive rate.
\end{abstract}




\maketitle

\section{Introduction}
\Ac{lbs} are integral to daily life, relying on positioning provided by terrestrial and satellite infrastructures, e.g., cellular network (3/4/5G), Wi-Fi, Bluetooth, and \ac{gnss} (e.g., \ac{gps}). Popular examples include navigation with Google Maps to a \ac{poi}, ride-hailing through Uber, and food delivery services. The correct position of the platform ensures the correct functionality of the application and the quality of the provided service. 

Recent real-world vulnerabilities of \ac{lbs} applications emerged \cite{PaaKjeIntTho:C18,Wan:J16,WanWanWanNik:J18,EryPap:C22,YilCukEmi:J23}, resulting even in scamming. Several attacks generate false position data fed into \ac{lbs}: players of location-based games (Pok{\'e}mon GO \cite{PaaKjeIntTho:C18}); or, scooter-sharing services and public transport, usually managed by geofencing or e-ticketing, virtually restricting the riding area or billing according to traveled distance \cite{YilCukEmi:J23,MicJatZibRaz:C24}. Position manipulation attacks can help win the game or break the geofencing, with the scooter seemingly within limits but in reality possibly far outside the fence, or allow traveling for free. Moreover, ``ghost drivers'' \cite{Wan:J16}, automatically assigned to passengers based on spoofed positions, never physically reach passengers but only pretend picking them up, yet charge fees---perpetrating a taxi fee scam. 

Position manipulation methods include \ac{gnss} spoofing, Wi-Fi spoofing, and \ac{vpn} proxies. \Ac{gnss} spoofers \cite{PsiHumSta:J16,SheWonCheChe:C20,AltMukKam:J23,YanEstVor:C23} broadcast adversarial satellite signals, either replayed or generated using open-source simulators \cite{LenSpaPap:C22}, transmitted with higher power but in the correct format fool receivers to lock on to them instead of the actual \ac{gnss} ones. Wi-Fi geolocation, increasingly relevant in urban environments and indoor settings, often assisting \ac{gnss}, can also be manipulated. Attackers broadcast pre-recorded or downloaded Wi-Fi beacons, using consumer-grade Wi-Fi routers or low-cost Wi-Fi chips \cite{TipRasPopCap:C09,VanPie:C14,Skylift2016,HanXioSheLu:C22,HanXioSheWei:J24}. Cellular-based positioning based on base station signals, can also be attacked, by replayed signals or deploying rogue base stations \cite{ShaBorParSei:C18}. IP geolocation methods (GeoIP) are also susceptible to manipulation, based on relays, transparent proxies, or \acpl{vpn} to control \ac{rtt} and positioning \cite{AbdMatVan:C17,KohDia:C22}. 

Highlighting the severity of \ac{lbs} position manipulation, this study examines the impact of joint attacks targeting \ac{gnss} and other wireless signals and implements specific attacks. Specifically, we manipulate position estimates from different infrastructures in a coordinated manner. As a special case, given Wi-Fi beacons are a weak point, we jam \ac{gnss} and replay (or forge) Wi-Fi beacons to control positioning results. 

Present solutions for \ac{gnss} spoofing detection, and more generally, secure positioning are mostly not designed for off-the-shelf platforms. They often rely on specialized hardware (e.g., resilient \ac{gnss} receivers or special-purpose antennas). For instance, a multi-antenna array is needed to calculate the \ac{aoa} for \ac{gnss} spoofing detection \cite{SchRadCamFoo:J16,BaiSunDemZha:J24}; or Wi-Fi \ac{csi} fingerprints require certain network interfaces to measure \cite{LinGaoLiDon:C20,YanYanYanSon:J22}. Many modern smartphones have no such antennas or modules resilient to interference \cite{Gps:J24}. Even so, recent proposals thwart attacks based on the assumption that some other infrastructure is out of reach of the adversary. For example, \cite{KasKhaAbdLee:J22,OliSciIbrDip:J22,LiuPap:C23} detect \ac{gnss} spoofing while assuming Wi-Fi or cellular signals are benign. 

In response to these challenges, our solution leverages diverse opportunistic ranging and motion data sources to detect position manipulation attacks. While previous work often assumes certain signals are benign, our approach considers the possibility that all wireless signaling used for positioning may be compromised, leading to manipulation and disruption of \ac{lbs} applications. By cross-validating opportunistic ranging information from \ac{gnss} and terrestrial network infrastructures, our approach is compatible and complements hardware fingerprinting or signal processing based attack detection \cite{Del:J24,LinGaoLiDon:C20}. Additionally, it independently improves detection accuracy and mitigates the impact of \ac{lbs} position manipulations from \ac{gnss} spoofing and rogue \acpl{ap}. 

We propose an extension of the \ac{raim} technique, working on multiple subsets of distance estimates from satellites and network signals, then cross-validating position estimations on subsets of distances with onboard sensors. As it is almost impossible to jam all benign signals, there are almost surely benign subsets that can be the basis for detecting the attack. Our proposed method includes two main phases. First, it leverages ranging information---distances derived from \ac{gnss} signals and terrestrial infrastructures such as Wi-Fi and cellular. Subsets of ranging information are generated to compute multiple intermediate position estimates, each associated with uncertainty. To improve the efficiency and accuracy of this process, we incorporate a subset sampling strategy. The second phase is position fusion with onboard sensors to detect and mitigate attacks. By cross-validating intermediate position estimates, our algorithm identifies inconsistencies indicative of position manipulation, such as \ac{gnss} spoofing or rogue \ac{ap} attacks. Unlike conventional detection methods that rely on a single infrastructure or unsecured fusion strategies, our scheme integrates diverse data sources in off-the-shelf platforms to improve robustness. 

Our contributions are: We demonstrate geolocation API attacks and illustrate how they can disrupt \ac{lbs} applications\footnote{\color{blue}{\url{https://drive.google.com/drive/folders/1rZtwVYXi3OwKyS8Yzn23E2E18rBP-J3x}}}. Building on these insights, we develop a \ac{raim}-based framework for detecting and mitigating threats in \ac{lbs} position manipulation. Different from \ac{raim} on \ac{gnss}, we use distance estimates from terrestrial networking infrastructures and \ac{gnss} with onboard sensors. By integrating this opportunistic ranging information from multiple sources, our approach gives an enhanced security likelihood against \ac{gnss} spoofing, rogue Wi-Fi \acpl{ap}, and other position manipulations. Our evaluation confirms the effectiveness of our approach, with improved true positive rate and delay in detecting attacks in various real-world scenarios. 

In the rest of the paper: Section~\ref{sec:bacgro} provides background knowledge of \ac{lbs}, \ac{gnss} spoofing, and Wi-Fi attacks. Section~\ref{sec:sysadv} presents our system model and adversary. Section~\ref{sec:attlbs} shows how to launch attacks on \ac{lbs} with details in the appendix. Section~\ref{sec:prosch} proposes our countermeasure. Section~\ref{sec:experi} evaluates the proposed scheme with baseline methods, and Section~\ref{sec:relwor} reviews related work about detection before we conclude in Section~\ref{sec:conclu}. 

\section{Background and Preliminaries}
\label{sec:bacgro}
\subsection{Location-Based Services}
\Ac{lbs} has reshaped industries ranging from navigation to marketing. Sensor fusion for precise and secure position estimation combines information from the \ac{imu}, \ac{lidar}, and network signals in vehicles or smartphones \cite{SheWonCheChe:C20,LiuPap:C23}. Indoor navigation systems bridge the gap between outdoor and indoor environments, addressing \ac{gnss} limitations in indoor settings, using technologies such as Wi-Fi, Bluetooth beacons, and \ac{uwb} ranging \cite{FurCriBarBar:J21}. Geofencing allows merchants to define virtual boundaries and trigger actions such as location-based messaging and dynamic pricing when users enter predefined areas \cite{RodDev:C14}. In food delivery, \ac{lbs} optimize shipping routes and reward participating taxis and couriers, enhancing efficiency \cite{LiuGuoCheDu:J18}. Enabling targeted advertising, location-based marketing strategies have redefined advertising, with emphasis on contextual offers boosting engagement and conversion rates \cite{LeWan:J20}. 

\subsection{GNSS Spoofing Attacks}
Spoofing \ac{gnss} involves transmitting fake but correctly formatted \ac{gnss} signals \cite{HumLedPsiOha:C08,SatStrLenRan:C22,SpaPap:C23}, to change position, timing, signal strength, and arrival angle, at the victim \ac{gnss} receiver, possibly hard to detect. Beyond efforts to authenticate \ac{gnss} signals and messages \cite{FerRijSecSim:J16,Gmv:J21b} to mitigate spoofing attacks, adversaries can still record and replay authentic \ac{gnss} signals \cite{MaiFraBluEis:C18,LenSpaPap:C22}. Authentication is not yet widely supported by receivers and needs extra computational overhead and modification. A more advanced replay/relay attack \cite{ZhaLarPap:J22} employs distance-decreasing attacks to tamper with signal timing, creating the false impression of earlier arrival. Recent research developed strategies to evade detection, such as slow variation to bypass tightly coupled \ac{gnss}/\ac{imu} systems \cite{SheWonCheChe:C20,GaoLi:J22} or gradual spoofing algorithms targeting \ac{gnss} time \cite{GaoLi:J23}. 

\subsection{Rogue Wi-Fi AP Spoofing}
Rogue Wi-Fi \acpl{ap} are unauthorized devices posing significant cybersecurity threats \cite{AloEll:J16}. These \acpl{ap} mimic legitimate \acpl{ap} and can intercept and alter client wireless communication, compromising integrity and confidentiality \cite{ThaRifGar:J22}. Wi-Fi attacks, including continuous or selective jamming and man-in-the-middle, have been demonstrated using commodity hardware in \cite{VanPie:C14}. Manipulating \acpl{rssi} and related positioning statistics by rogue \acpl{ap} introduces position inaccuracies and inconsistencies \cite{FluPotPapHub:C10,YuaHuLiZha:C18,HanXioSheWei:J24}. Open-source tools for broadcasting Wi-Fi beacons can manipulate smartphone positioning results \cite{Skylift2016}. Furthermore, rogue \acpl{ap} deceiving Wi-Fi clients to automatically connect \cite{LouPerSta:J23}, can delay, relay, or replay client packets, thereby manipulating GeoIP positioning results \cite{AbdMatVan:C17,KohDia:C22}. 

\section{System Model and Adversary}
\label{sec:sysadv}
\subsection{System Model}
\begin{figure}
\begin{centering}
\includegraphics[trim={0 0 2cm 0},clip,width=\columnwidth]{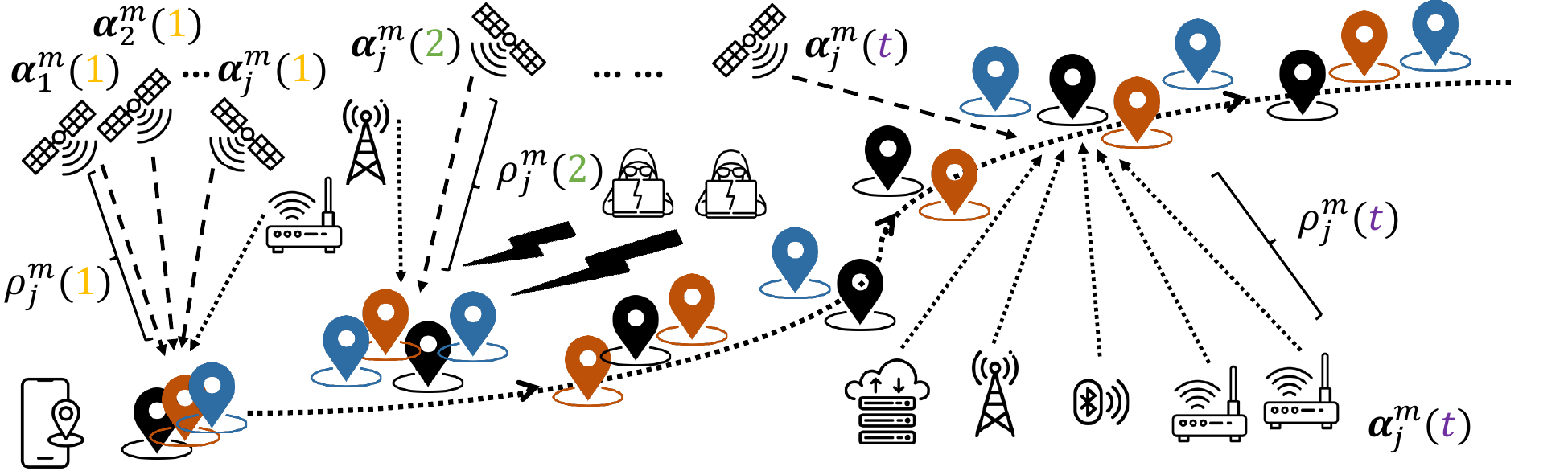}
\par\end{centering}
\caption{\Ac{lbs} applications have ranging information with anchor positions from \ac{gnss} and network infrastructures.}
\label{fig:sysmod}
\end{figure}

As shown in Figure~\ref{fig:sysmod}, the system considers \ac{lbs} deployed on mobile platforms (e.g., smartphone, tablet, or intelligent vehicle) that can process both \ac{gnss} and other opportunistic signals to position themselves (the devices). At time $t$, the true position of the platform, denoted as $\mathbf{p}_{\text{usr}}(t) \in \mathbb{R}^3$, is estimated by using \ac{gnss} and opportunistic information to compute $\mathbf{p}_{\text{lbs}}(t)$. Opportunistic information includes wireless signals received from network modules (e.g., Wi-Fi, cellular networks), Bluetooth, GeoIP, as well as motion data from onboard sensors (e.g., \acpl{imu} or wheel speed sensors). 

Position estimates from different sources have varying accuracy; however, in benign conditions without any attacks, these estimates are expected to be consistent. If position manipulation occurs, significant deviations between $\mathbf{p}_{\text{lbs}}(t)$ and $\mathbf{p}_{\text{usr}}(t)$ are expected. 

\textbf{Notation:} Denote motion measurements as velocity, $\mathbf{v}(t)$, acceleration, $\mathbf{a}(t)$, and orientation, $\boldsymbol{\omega}(t)$. Ranging information, denoted as $\rho_j^m(t)$, is associated with the positions of anchors $\boldsymbol{\alpha}_j^m(t)$; these are \ac{gnss} satellites, cellular \acpl{bs}, Wi-Fi \acpl{ap}, Bluetooth devices, and GeoIP \ac{rtt} servers; $j \in \mathcal{J}^m(t)$, where $\mathcal{J}^m(t)$ is the set of anchors at time $t$, $m=1,2,...,M$, and $M$ is the number of opportunistic information sources. 

\begin{figure}
\begin{centering}
\includegraphics[width=\columnwidth]{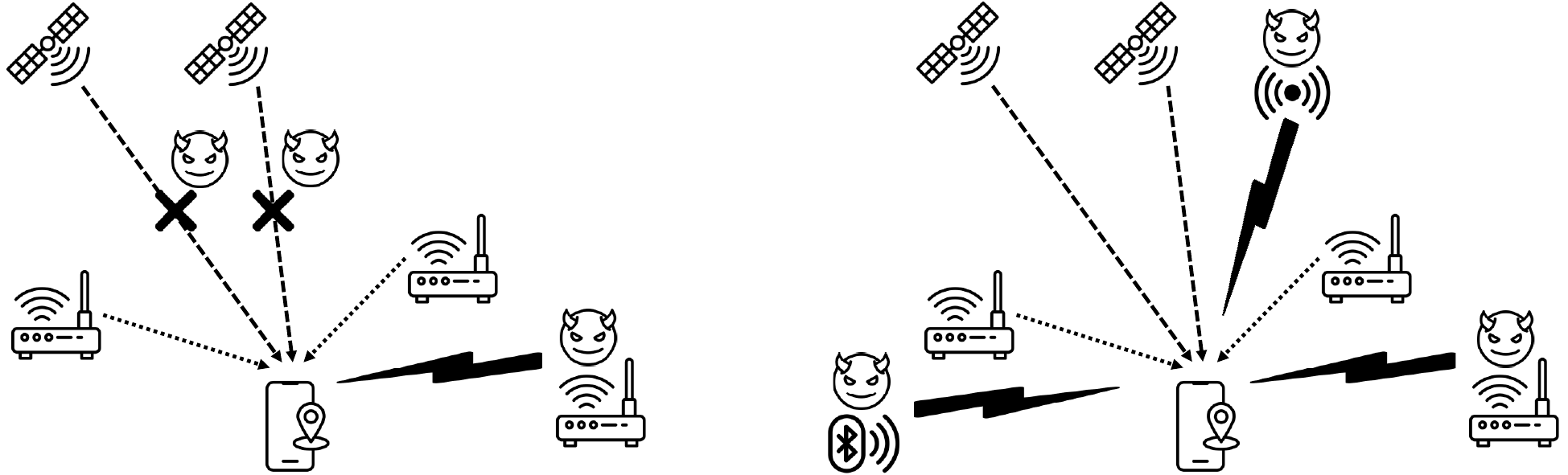}
\par\end{centering}
\caption{Left: Wi-Fi spoofing attack with \ac{gnss} jamming; right: coordinated location spoofing.}
\label{fig:advers}
\end{figure}

\subsection{Adversary Model}
The adversary knows $\mathbf{p}_{\text{usr}}(t)$ and aims to control $\mathbf{p}_{\text{lbs}}(t)$, to disrupt the integrity and security of \ac{lbs}. By manipulating wireless signals (e.g., \ac{gnss} pseudorandom noise codes, Wi-Fi beacons, cellular signals, Bluetooth beacons) and messages (e.g., \ac{rtt}-related packets), it can compromise ranging information $\rho_j^m(t)$ used for positioning. We assume all wireless signals can be potentially attacked, although realistically, the adversary manipulates a subset of them at any given time. The following attack types are considered. 

\subsubsection{Wi-Fi Spoofing with GNSS Jamming} As \ac{gnss} spoofing attacks are feasible but still require a degree of sophistication, we propose jamming \ac{gnss} persistently, a relatively easy task, and force \ac{lbs} to rely on network-based positioning information. The adversary injects falsified ranging information only by generating and replaying Wi-Fi beacons captured from another place (as illustrated in the left part of Figure~\ref{fig:advers}). These attack signals persist throughout the entire attack lifecycle but coexist with some legitimate transmissions, including original Wi-Fi and cellular signals.

\subsubsection{Coordinated Location Spoofing} The adversary coordinates a sequence of falsified positions for the victim. It spoofs \ac{gnss} and replays cellular, Wi-Fi, and Bluetooth (and forges some of them, if possible), along the predetermined trace the attacker wishes to mislead the victim into perceiving. Through gradual deviation, the victim believes it is following a legitimate path. We assume that attack signals coexist with benign signals. Although the adversary may selectively jam Wi-Fi or cellular communication, it is infeasible to eliminate all legitimate transmissions at all times; it could also cause outages that would be easily detected.

\section{Attack Demonstrations}
\label{sec:attlbs}
We demonstrate successful \ac{lbs} position manipulation attacks based on the adversary model. All demonstrations were conducted with full attention to ethical considerations, in controlled environments with absolutely no impact on actual systems and users.

\subsection{Multi-Band GNSS Spoofing}
We demonstrate how a multi-band \ac{gnss} spoofing attack can manipulate fused location providers, such as Google Maps, with the attack applicable to other applications and services. The attack steps with detailed experimental configurations in Section~\ref{subsec:expsetb} are: 

\textbf{Initial \ac{gnss} Jamming:} Before spoofing, \ac{gnss} jamming forces the victim receiver to lose its lock on authentic signals. \textbf{Spoofing Signal Generation:} Using a \ac{gnss} signal generator (Skydel with USRP N310), the attacker broadcasts spoofing signals on multiple constellations and frequency bands (\ac{gps} L1/L5 and Galileo E1/E5a). These signals are transmitted at a higher power level than the real ones. \textbf{Dominant Position Estimation:} Although the smartphone location provider fuses data from \ac{gnss}, Wi-Fi, cellular, and Bluetooth, our attack ensures spoofing \ac{gnss} signals exhibit a favorable \ac{dop} to dominate the fusion algorithm. 

\textbf{Attack Results:} Although the \ac{gnss} position is not consistent with the network-based positioning result, \ac{lbs} applications directly provide the fused position near the spoofing position, thus imposing attacker control on the victim. If the user shares location data, the spoofed \ac{gnss} position will be used to train and build the network positioning and geolocation database of \ac{lbs} providers. Note that attacking participatory sensing (e.g., on Google Maps \cite{EryPap:C22}) can be done without physical presence and attack.

\subsection{Wi-Fi Spoofing with GNSS Jamming}
\label{subsec:attlbsdeg}
To overcome the high cost and portability issues of \ac{gnss} spoofers, we design a low-cost and easily deployable attack that leverages the replay of Wi-Fi beacons while in parallel jamming \ac{gnss}: 

\textbf{Enduring \ac{gnss} Jamming:} Rather than a one-time jamming event, the adversary maintains \ac{gnss} jamming throughout the attack to prevent the reception of authentic \ac{gnss} signals. \textbf{Wi-Fi Beacon Replay:} A commercial Wi-Fi router (Linksys WRT1200AC) replays pre-recorded or artificially generated Wi-Fi beacons mapped to a predetermined spoofing position or trace. Unlike \ac{gnss} signals, Wi-Fi beacons do not require precise time synchronization, making them easier to manipulate. \textbf{Fallback Positioning:} With \ac{gnss} signals blocked, modern mobile devices rely on network-based positioning as a fallback. The replayed beacons thus mislead the fusion algorithm, causing it to compute a spoofed position. 

\begin{figure}
\includegraphics[trim={0 5cm 0.1cm 3.2cm},clip,width=.515\columnwidth]{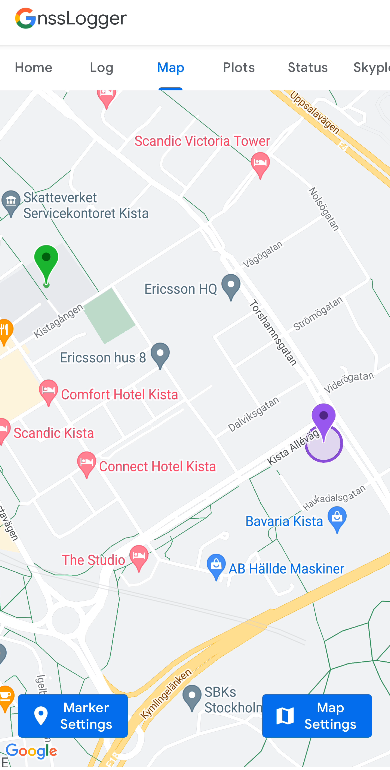}
\hfill
\includegraphics[trim={0.8cm 7.5cm 0.1cm 6.5cm},clip,width=.45\columnwidth]{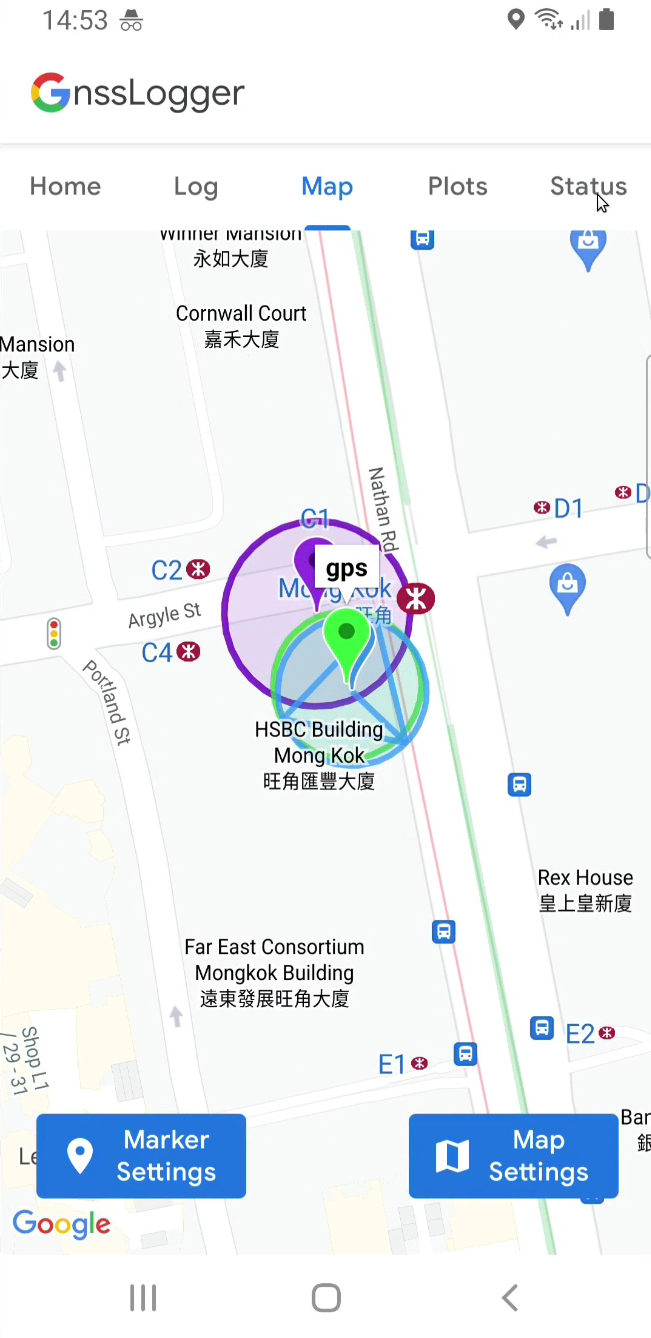}
\caption{Left: Wi-Fi spoofing attack with \ac{gnss} jamming manipulates the application in the upper left corner (dark green pin) to the position at the bottom right (purple pin). Right: coordinated spoofing attack manipulates the network-based (purple circle with pin), \ac{gnss} (light green circle with pin), and fused (blue circle with pin) positions, deviating from Europe to Hong Kong.}
\label{fig:degcor}
\end{figure}

\textbf{Attack Results:} We observe that both Android and iOS devices fail to notify users when positioning relies solely on (possibly adversarial) network signals, potentially exposing users to undetected \ac{lbs} position manipulation. For instance, a taxi driver may falsify the driving route to get illegal profits or bypass trip security checking to take the passenger to an unintended destination without any notification. Figure~\ref{fig:degcor} (left) and Appendix~\ref{app:appenda} illustrate the effectiveness of the position manipulation. 

\subsection{Coordinated Location Spoofing}
\label{subsec:attlbscor}
Security-sensitive applications, such as mobile banking, typically validate location data by cross-referencing multiple sources (e.g., \ac{gnss}, Wi-Fi, and IP address). If the device meets verification criteria, region lock for the usage is lifted. Mobile payments, such as Revolut Card and WeChat Pay, have location-based security, comparing the mobile device position with the place of offline payment. If the two positions deviate, the transaction may be declined. To bypass them, as an example, given the attack is of broader interest, we designed a coordinated spoofing attack that manipulates all positioning. 

\textbf{Simultaneous Signal Manipulation:} The attacker uses both a \ac{gnss} signal generator and a Wi-Fi router to broadcast spoofing signals to manipulate \ac{gnss} and network positions consistent with the spoofing position. \textbf{Network Beacon Crafting:} Wi-Fi beacons are replayed by using data extracted from public databases \cite{BobArkUht:J23}. Since geolocation APIs do not verify the authenticity of beacons, the attacker can generate them without physical presence at the spoofing position. \textbf{Network Traffic Redirection:} The router relays all TCP and UDP packets from the connected devices to a cloud server located near the intended spoofing position using iptables. 

\textbf{Attack Results:} By ensuring that the spoofing of \ac{gnss}, Wi-Fi, and (if applicable) Bluetooth signals all point to the same position, as shown in Figure~\ref{fig:degcor} (right), our coordinated attack successfully bypasses cross-validation mechanisms deployed by secure applications. The detailed demonstrations are available in Appendix~\ref{app:appendb}. 

\begin{algorithm}%[hbt!]
\begin{flushleft}
\hspace*{\algorithmicindent} \textbf{Input} $\{\boldsymbol{\alpha}_j^m(t), \rho_j^m(t), {\mathbf{v}}(t), {\mathbf{a}}(t), \boldsymbol{\omega}(t)\}, \mathbf{p}_{\text{lbs}}(t)$\\
\hspace*{\algorithmicindent} \textbf{Parameter} $\Lambda_f$\\
\hspace*{\algorithmicindent} \textbf{Output} \textit{AttackDetected}
\end{flushleft}
\begin{algorithmic}[1]
\For{$t = 1$, $t{+}{+}$} \Comment{Time index}
\For{$m = 1$, $m{+}{+}$, $m \le M$} \Comment{Infrastructures}
\For{$l = 1$, $l{+}{+}$, $l \le L^m(t)$} \Comment{All subsets}
\State $\mathbf{p}_l^m(t) \gets$ Section~\ref{subsec:posmet} \Comment{Positioning}
\State $\hat{\mathbf{p}}_l^m(t) \gets$  Section~\ref{subsec:motinf} \Comment{Smoothing}
\State $\hat{\boldsymbol{\sigma}}_l^m(t) \gets$ Section~\ref{subsec:uncmod} \Comment{Uncertainties}
\State $f_{l,t}^m(\mathbf{p})={\frac {1}{\hat{\boldsymbol{\sigma}}_l^m(t) {\sqrt {2\pi }}}}\exp \left(-{\frac {1}{2}}\left({\frac{\mathbf{p} - \hat{\mathbf{p}}_l^m(t)}{\hat{\boldsymbol{\sigma}}_l^m(t)}}\right)^2\right)$
\Statex \Comment{Probability density function}
\EndFor
\EndFor
\State $f_t(\mathbf{p})=1-\left(\prod_{m=1}^{M}\left(\prod_{l=1}^{L^m(t)} f_{l,t}^m(\mathbf{p})\right)^{\frac{1}{L^m(t)}}\right)^{\frac{1}{M}}$
\Statex \Comment{Likelihood function and fusion}
\If{$f_t(\mathbf{p}_{\text{lbs}}(t)) < \Lambda_f$} 
    \State \textit{AttackDetected} $=$ \textit{True}
\Else
    \State \textit{AttackDetected} $=$ \textit{False}
\EndIf
\EndFor
\end{algorithmic}
\caption{Detection based on subset generation and cross-validation using available opportunistic information}
\label{alg:extraim}
\end{algorithm}

\section{Defense Scheme}
\label{sec:prosch}
We propose a scheme easily deployable on mobile devices, using readily available opportunistic information, without assuming the presence of a trusted location source, considering that attackers can target multiple positioning methods simultaneously. Our scheme extends the \ac{raim} method by leveraging redundant and opportunistic information to detect \ac{lbs} position manipulation. It integrates multiple information sources, which include \ac{gnss}, \ac{sop} (Wi-Fi, cellular, Bluetooth, GeoIP, etc.), and motion data (velocity, acceleration, and orientation based on on-board sensing), to strengthen detection. The overall process consists of two main steps: subset generation for positioning and position fusion for attack detection, as outlined in Algorithm~\ref{alg:extraim}.

First, real-time ranging information is collected from all available infrastructures. Through strategically combining different ranging information from heterogeneous infrastructures into subsets, it accommodates variations in signal characteristics, such as distance, uncertainty, and accuracy. As it is almost impossible to jam all benign anchors, it is reasonable to expect some benign subsets exist. For example, Wi-Fi jamming \cite{VanPie:C14} cannot jam all beacons, as Wi-Fi channels are wide and relatively resistant to interference. To reduce computational complexity, as subset sizes range from the minimum required by the positioning algorithm to the maximum, our sampling strategy (Section~\ref{subsec:subran}) reduces the number of subsets and improves efficiency. Finally, intermediate position estimates are computed for each subset and infrastructure, using the corresponding positioning algorithms in Section~\ref{subsec:posmet}. 

Second, onboard sensors collect velocity, acceleration, and orientation data to refine intermediate position estimates obtained in the first step. The proposed local polynomial regression in Section~\ref{subsec:motinf} smooths positions based on the physical movement constraints of the mobile platform. The filtered positions are then fused into probability density functions along with their associated uncertainties, as described in Section~\ref{subsec:uncmod}. A composite function in Section~\ref{subsec:likfun} is normalized to derive the final likelihood used for cross-validating position manipulations. 

\subsection{Subset Generation}
\label{subsec:subgen}
\subsubsection{Raw Data Preprocessing}
The input data is the same crowdsourced data used for \ac{lbs} applications. Information from \ac{gnss} satellites, Wi-Fi \acpl{ap}, cellular \acpl{bs}, Bluetooth devices, GeoIP \ac{rtt} servers, and onboard sensors is recorded as they are available. \Ac{gnss} signal data includes received times, satellite positions, and pseudoranges (estimates of satellite-receiver distances). Wi-Fi beacons include \ac{bssid}, \ac{ssid}, and \ac{rssi}. Cellular data includes cell identifier, and \ac{rssi}. Bluetooth beacons include \ac{mac} and \ac{rssi}. GeoIP data includes the IP address of the platform and \ac{icmp} messages containing \ac{rtt} like ping values to data centers around the world. Onboard sensors provide motion measurements. All this information is timestamped and different types of data are temporally aligned. 

\ac{gnss} pseudoranges, constant-biased approximation of the distance between the satellite and the receiver, are calculated by the time it takes for the signal to reach the receiver, multiplied by the speed of light. The pseudorange error may accumulate to hundreds of meters after a few seconds because the receiver clock quartz oscillator drifts. The clock is used to measure pseudoranges; thus, all pseudoranges have the same clock error factor. Then, in a benign environment, this ranging information can still accurately position the receiver by adopting at least four satellites, solving for its coordinates and clock error. Under a jamming attack, pseudoranges cannot be derived, partially or entirely. Under spoofing, pseudoranges of one or more constellations are modified by the attacker, i.e., the derived ranging information deviates from the benign one. 

For Wi-Fi, cellular, and Bluetooth, the received opportunistic ranging information does not represent actual distances. Instead, it is \ac{rssi}, a negative number in dBm, which follows a log-distance path loss model. The closer the device, the stronger the signal, and vice versa. This ranging information is a relatively inaccurate approximation of distance compared to \ac{gnss}. Under jamming, the platform cannot receive valid packets, thus \ac{rssi} cannot be derived.  

For GeoIP, both the IP address and \ac{icmp} based ranging are used. \Ac{icmp} utilities (e.g., ping and traceroute) provide time delay (e.g., \ac{rtt}). The delay multiplied by half the speed of light should be an approximation of the distance between the platform and the GeoIP \ac{rtt} server. In the situation of delay manipulation attacks at the physical and data link layers (e.g., Wireshark), \ac{icmp} messages forwarded cause delay, thus longer ranges. In the situation of attacks using transparent proxy, this proxy is usually running on the network layer, so only TCP and UDP messages are managed rather than \ac{icmp} messages. Firewalls may be used to drop/reject \ac{icmp} messages, analogously to jamming, thus preventing ranging information from ping values.

Most importantly, we have a database of anchors, containing positions of Wi-Fi \acpl{ap}, cellular \acpl{bs}, Bluetooth devices, and GeoIP servers, playing a role analogous to \ac{gps} ephemerides. Among all the anchors in the database, data cleaning can eliminate a majority of incorrect or non-fixed anchors, e.g., personal hotspots, Bluetooth headphones, and public transport Wi-Fi \acpl{ap}. 

\subsubsection{Subsets of Ranging Data}
\label{subsec:subran}
Subsets are generated to explore all possible combinations of anchors/constellations, from the minimum size required by positioning to the maximum. This process does not assume a specific number of attacked ranging information sources (e.g., spoofed constellations, rogue \acpl{ap}, and delayed \acpl{rtt}), thus being applicable to any potential scenario. 

The \ac{gnss} subsets include all possible combinations of constellations, including \ac{gps}, Galileo, GLONASS, and Beidou. For \acpl{ap}, \acpl{bs}, or Bluetooth anchors, the receiver position can be determined using at least three \acpl{rssi} in trilateration, and the clock error cannot be estimated. Similarly, for GeoIP, at least three \acpl{rtt} to determine the rough position. Hence, their number of subsets is $\sum_{i=3}^{J^m(t)}C(J^m(t),i)$, where $J^m(t)=|\mathcal{J}^m(t)|,\forall m>1$. These subsets of ranging information (associated with anchor) indexes, $j$ (from $\rho_j^m(t)$), are denoted as $\mathcal{S}_l^m(t)$, where $l=1,2,...,L^m(t)$, with $L^m(t)$ the total number of subsets for the $m$th infrastructure (\ac{gnss}, Wi-Fi, etc.). Then, for each $(l,m)$, we use $\mathbf{p}_l^m(t)$ as the subset positioning result based on $\mathcal{S}_l^m(t)$.

The number of generated subsets for localization may be very large, leading to sizable computational complexity; therefore, we use a subset sampling strategy: randomly select subsets before the next positioning step, with every subset selected or not via a predetermined probability distribution. For example, a discrete uniform distribution makes subsets equally likely to be chosen, not introducing bias or skew to $\hat{\mathbf{p}}_{\text{usr}}(t)$, thus does not undermine the cross-validation process. Through randomly choosing subsets, our detection scheme stays robust and adaptable to heterogeneous opportunistic information sources and attack types. Most significantly, it reduces the detection computational complexity. 

\subsubsection{Positioning Methods}
\label{subsec:posmet}
In order to use the heterogeneous ranging information provided by multiple infrastructures, we have off-the-shelf positioning methods for the subsets from each data type, e.g., trilateration, multilateration, and geolocation localization. Each provides position estimation with an uncertainty value. 

Trilateration for \ac{gnss} single point positioning is based on code observations of pseudoranges. The observations are affected by errors such as atmospheric delay, satellite clock, and receiver clock errors. GLONASS and \ac{gps} have differences in the way the ionospheric and tropospheric delays are modeled. Additionally, GLONASS uses a different frequency band than \ac{gps}, so the wavelength of the carrier wave is different. The pseudorange between the $j$th satellite and a user at $\mathbf{p}_{\text{usr}}(t)$ is $\rho_j^m(t)=||\mathbf{p}_{\text{usr}}(t)-\boldsymbol{\alpha}_j^m(t)||+\epsilon_\text{n}$, where $\epsilon_\text{n}$ models errors. Then, positioning uses the pseudoranges between the receiver and the satellite positions to compute the receiver position: $||\hat{\mathbf{p}}_{\text{usr}}(t)-\boldsymbol{\alpha}_j^m(t)||=\rho_j^m(t), j \in \mathcal{S}_l^m(t)$. 

Geolocation is distance-based positioning based on a weighted least squares problem to minimize the weighted sum of squared distances between anchors and the estimated device/user position \cite{Mozilla2023}. These weights are determined based on the inverse square of ranging: $\underset{\hat{\mathbf{p}}_{\text{usr}}(t)}{\mathop{\min}}\quad \sum_{j \in \mathcal{S}_l^m(t)} \left( {||\hat{\mathbf{p}}_{\text{usr}}(t)-\boldsymbol{\alpha}_j^m(t)||}/{\rho_j^m(t)} \right)^2$, solved by numeric minimization algorithms such as the SciPy least squares. 

GeoIP positioning combines both tabulation-based and delay-based IP geolocation \cite{AbdMatVan:C17}. Tabulation-based IP geolocation provides a lookup table to map IP address to an estimated position. Delay-based IP geolocation uses \acpl{rtt} as ranging information with 10 to 20 anchors. It first maps \ac{rtt} to distance based on a fitted function from training data. Then, the position is estimated as the centroid of the intersection of circles whose centers are the anchors and radii are the distances. 

\subsection{Position Fusion}
\label{subsec:locfus}
\subsubsection{Onboard Sensors}
\label{subsec:motinf}
We use $\mathbf{p}_l^m(t)$ from Section~\ref{subsec:subgen} and ${\mathbf{v}}(t)$, ${\mathbf{a}}(t)$, $\boldsymbol{\omega}(t)$ from onboard sensors as input data, applying local polynomial regression to filter noise. Positions, $\mathbf{p}_{\text{usr}}(t), \allowbreak\mathbf{p}_{\text{lbs}}(t), \mathbf{p}_l^m(t)$, are represented in the format of the World Geodetic System 1984 (WGS84). $\boldsymbol{\omega}(t) \in \mathbb{R}^3$, from onboard sensors, comprises roll ($\phi$), pitch ($\theta$), and yaw ($\psi$) angles in the format of the sensor coordinate system of the mobile platform. The rotation matrix, $\mathbf{R}$, below converts a smartphone's sensor coordinates to WGS84:
\begin{multline*}
\mathbf{R}(t) =\mathbf{R}_{\phi}(t)\mathbf{R}_{\theta}(t)\mathbf{R}_{\psi}(t) =
\left[\begin{array}{ccc}
\cos\psi(t) & -\sin\psi(t) & 0\\
\sin\psi(t) & \cos\psi(t) & 0\\
0 & 0 & 1
\end{array}\right]\\
\qquad\times 
\left[\begin{array}{ccc}
1 & 0 & 0\\
0 & \cos\theta(t) & \sin\theta(t)\\
0 & -\sin\theta(t) & \cos\theta(t)
\end{array}\right]
\left[\begin{array}{ccc}
\cos\phi(t) & 0 & -\sin\phi(t)\\
0 & 1 & 0\\
\sin\phi(t) & 0 & \cos\phi(t)
\end{array}\right].
\end{multline*}
This definition differs from that used in aviation, with $\phi$ and $\theta$ interchanged, and $\psi$ changes the direction of rotation. The state of the mobile platform, $\big(\mathbf{p}_{\text{usr}}(t),\mathbf{v}(t),\mathbf{a}(t) \big)$, evolves over time, thus from $t-1$ to $t$, so
\begin{align*}
\mathbf{p}_{\text{usr}}(t)&=\mathbf{p}_{\text{usr}}(t-1)+\mathbf{R}(t-1)\mathbf{v}(t-1)+\frac{1}{2}\mathbf{R}(t-1)\mathbf{a}(t-1)+\mathbf {n}\\
\mathbf{v}(t)&=\mathbf{v}(t-1)+\mathbf{a}(t-1)+\mathbf {n}
\end{align*}
where $\mathbf {n}$ models noise. The state transition matrix is
\begin{equation}
    \mathbf {F}(t) ={\begin{bmatrix}\mathbf{1}&\mathbf{R}(t)\\\mathbf{0}&\mathbf{1}\end{bmatrix}}
\end{equation}
and the control-input matrix is $\mathbf {B}(t) =\begin{bmatrix}\mathbf{\frac{1}{2}}\mathbf{R}(t) & \mathbf{1}\end{bmatrix} ^\text{T}$. We denote the estimated state after movement as 
\begin{equation}
    \begin{bmatrix}\overline{\mathbf{p}}_l^m(t)\\\overline{\mathbf{v}}(t)\end{bmatrix}=\mathbf {F}(t-1) \cdot \begin{bmatrix}\mathbf{p}_l^m(t-1)\\\mathbf{v}(t-1)\end{bmatrix}+\mathbf{B}(t-1) \cdot \mathbf{a}(t-1)
    \label{eq:eststamot}
\end{equation}
where $l=1,2,...,L^m(t), m=1,2,...,M$. Then, to combine the motion with regression, we use an estimator $\hat{\mathbf{p}}_l^m(t) = \mathbf{W}\mathbf{t}$, where $\mathbf{W} \in \mathbb{R} ^{3 \times (n+1)}$ is a matrix of polynomial coefficients, $n$ is the order of the polynomial regression, $\mathbf{t}$ is a $(n+1)$ dimensional vector, $[\mathbf{t}]_i=t^{i-1}$, and $\mathbf{W}$ at $(l,m,t)$ is calculated from the local polynomial regression problem:
\begin{equation}
    \begin{array}{*{20}{c}}
    {\mathop {\min }\limits_{\mathbf{W}} }&{\sum\limits_{t'=t-w}^{t} [\mathbf{W} \mathbf{t'}-\mathbf{p}_l^m(t')]^\top K_\text{loc}(t-t')[\mathbf{W} \mathbf{t'}-\mathbf{p}_l^m(t')]} \\ 
    {\text{s.t.}}&{|\mathbf{W} \mathbf{t} - \overline{\mathbf{p}}_l^m(t)| \le \boldsymbol{\epsilon_\text{t}}}
    \end{array}
\label{eq:proall}
\end{equation}
where $w$ is a rolling window of the filter and $K_\text{loc}(t-t')$ is a kernel function assigning a scalar value to ensure the closer data has a higher weight, e.g., $\exp \left(-(t-t')^2\right)$. $\boldsymbol{\epsilon_\text{t}} \in \mathbb{R}^3$ in the constraint is a small tolerance for $\overline{\mathbf{p}}_l^m(t)$ to ensure the estimated position satisfies the movement in a short time duration. $\mathbf{p}_l^m(t')$ may not be available during the whole window $w$, then we use $\hat{\mathbf{p}}_{\text{usr}}(t')$ to complement missing positions. 

The second derivative of the objective function in \eqref{eq:proall} with respect to $\mathbf{W}$ is $2 \cdot \sum\limits_{t'=t-w}^{t}  K_\text{loc}(t-t')\cdot (\mathbf{t}\cdot \mathbf{t}^\top )^\top \otimes \mathbb{I}$, where $\otimes$ is the Kronecker product. This derivative is always a positive definite matrix. In addition, the constraints are affine functions, so the problem is convex and solvable in polynomial time using Lagrange multipliers. After solving the problem, we have $\hat{\mathbf{p}}_l^m(t)$.

\subsubsection{Uncertainty Modelling}
\label{subsec:uncmod}
The positioning uncertainty differs across various infrastructures and positioning methods. For \ac{gnss}-based trilateration, we use the position \ac{dop} metric. $(\sigma_x,\sigma_y,\sigma_z,\sigma_t) = \text{DOP} \triangleq \sqrt{\text{Tr}(\mathbf{Q})}$, where $\mathbf{Q}$ is the covariance matrix of the least squares solution to the navigation equations and \ac{dop} is the square root of the trace of $\mathbf{Q}$. Then, the uncertainty $\hat{\boldsymbol{\sigma}}_l^m (t)$ is $(\sigma_x,\sigma_y,\sigma_z)$. For geolocation and other least squares algorithms, the uncertainty is represented by the residual of least squares. For other positioning techniques, we use the residual vector from the local polynomial regression to model the uncertainty of estimated positions.

\subsubsection{Likelihood Function}
\label{subsec:likfun}
The probability density function of each position estimate is defined as follows:
\begin{equation}
    f_{l,t}^m(\mathbf{p})={\frac {1}{\hat{\boldsymbol{\sigma}}_l^m(t) {\sqrt {2\pi }}}}\exp \left(-{\frac {1}{2}}\left({\frac{\mathbf{p} - \hat{\mathbf{p}}_l^m(t)}{\hat{\boldsymbol{\sigma}}_l^m(t)}}\right)^2\right)
\end{equation}
where where $\hat{\mathbf{p}}_l^m(t)$ and $\hat{\boldsymbol{\sigma}}_l^m(t)$ denote the estimated mean position and standard deviation of the $l$-th estimate from the $m$th infrastructure, and all the operations are element-wise. Then, to aggregate intermediate position estimates along with their uncertainties, assumed to follow distributions $\mathcal{N}(\hat{\mathbf{p}}_l^m(t), \hat{\boldsymbol{\sigma}}_l^m(t)^2)$, we define a composite likelihood function. At time $t$, the likelihood function of $\mathbf{p}_{\text{lbs}}(t)$ under attack is computed as 
\begin{equation}
    f_t(\mathbf{p}_{\text{lbs}}(t))=1-\left(\prod_{m=1}^{M}\left(\prod_{l=1}^{L^m(t)} f_{l,t}^m(\mathbf{p}_{\text{lbs}}(t))\right)^{\frac{1}{L^m(t)}}\right)^{\frac{1}{M}}
\end{equation}
where the cumulative term penalizes disagreement between $\mathbf{p}_{\text{lbs}}(t)$ and the fused distribution, yielding a higher likelihood when inconsistencies are detected. To determine whether position manipulation occurs, a threshold $\Lambda_f$ is predefined based on established detection metrics. For example, the Z‐score method sets the threshold by calculating the mean and standard deviation, while kernel density estimation non-parametrically estimates the probability density function to support threshold selection. We use \ac{roc} to analyze detection accuracy versus different false alarm rates in our evaluation results to assess the trade-off between them. If $f_t(\mathbf{p}_{\text{lbs}}(t))$ is larger than $\Lambda_f$, $\mathbf{p}_{\text{lbs}}$ is deemed the result of an attack. 

\section{Evaluation}
\label{sec:experi}
We conducted the experiments in two settings and collected two distinct datasets to evaluate our detection approach. Jammertest 2024 \cite{Jam:J24} provides real-world attacks (fixed position, dynamic, time jumping, etc.) solely on \ac{gnss}, but still allows for interesting results as the \ac{gnss} attacks affect other data. Coordinated location spoofing is done in a lab environment with more flexibility and control on the simulated attack strategies on \ac{gnss} and other positioning. 

\begin{figure}
    \centering
    \includegraphics[width=\columnwidth]{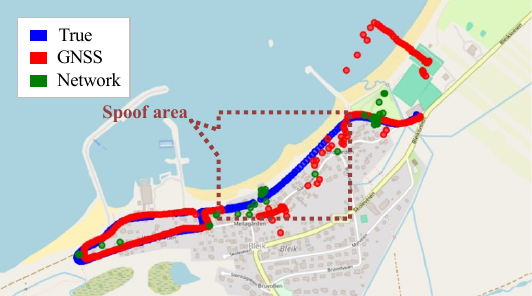}
    \caption{A driving trace, \ac{gnss} positions, and network-based positions. The red dotted line frames the attacked area. The network signals are not specifically attacked.}
    \label{fig:tracea}
\end{figure}

\subsection{Dataset A: Jammertest 2024}
To assess the real-world applicability of the proposed detection method against \ac{lbs} position attacks, we collect data in an open-air environment. This setting provides a rare opportunity to legally observe a variety of \ac{gnss} attacks alongside benign wireless signals.
\subsubsection{Experimental Setup}
The test environment is an outdoor area in Bleik and its surroundings, with intermittent \ac{gnss} jamming and spoofing by the organizers of Jammertest \cite{Jam:J24}. The attack equipment involved various types of jammers, meaconers, and spoofers. Cigarette jammers, handheld jammers, and fixed jammers targeted \ac{gnss} as well as mobile communication bands (GSM and DCS). In the high power \ac{gnss} jamming scenario, jamming-to-signal ratio exceeded 24 dB over a distance of up to 73 km. Following successful \ac{gnss} jamming, \ac{gnss} spoofing was initiated. The fixed meaconer was deployed on a mountain, retransmitting live-sky signals from a long fiber-optic cable. The spoofing included stationary spoofing of small/large position jumps, simulated driving, flying spoofing, and more, employing Skydel with two USRP X300 \acpl{sdr} to generate the \ac{gnss} signals following the pre-planned routes \cite{Jam:J24}. In most of the test cases, the location services on the smartphones were successfully deceived, as expected, with the position effectively manipulated. However, as shown in Figure~\ref{fig:tracea}, due to the intermittent nature of spoofing, signal blockage, environmental dynamics, and spoofing signals being mostly weaker than jamming, not all \ac{gnss} positions are spoofed. 

\subsubsection{Dataset Collection}
We collected 68 driving traces using Android smartphones, including the Samsung S9, Redmi 9, Google Pixel 4 XL, and Pixel 8 (covering chipsets from Exynos, MTK, Qualcomm, and Google Tensor). Additionally, two u-blox receivers (ZED-F9P) served for ground truth positioning. The spoofing attacks did not target GLONASS, Beidou, and QZSS constellations, thus the ZED-F9P receivers were set up with clear views of these unaffected \ac{gnss} constellations, as well as an external \ac{gnss} reference station outside the affected region to ensure precise kinematic positioning. 

Smartphones in the car captured potentially compromised \ac{gnss} and network signals. The GNSSLogger application recorded \ac{gnss} traces of the phone at a sampling rate of 1 Hz, consisting of RINEX and raw text files of satellite status with pseudoranges. The NetworkSurvey application recorded beacons and other messages from cellular, Wi-Fi, and Bluetooth anchors along the trace at approximately 0.3 Hz, in GeoPackage format, providing anchor names with \acpl{rssi}. In addition, acceleration (m/s$^2$), orientation (degrees), angular velocity (°/s), and magnetic field (mT), are provided at 100 Hz from onboard sensors. Satellite positions were obtained from the broadcast ephemeris, and the positions of \acpl{bs}, \acpl{ap}, and Bluetooth anchors were retrieved from a crowdsourced geolocation database calculated by the WiGLE.net application \cite{BobArkUht:J23}, based on both benign signals and measurements affected by adversaries or inadvertent contributors under \ac{gnss} attack. 

\begin{figure}
    \centering
    \includegraphics[width=.44\columnwidth]{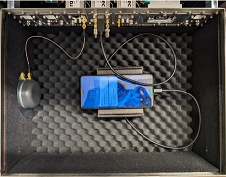}
    \hfill
    \includegraphics[trim={0 1.56cm 0 0},clip,width=.53\columnwidth]{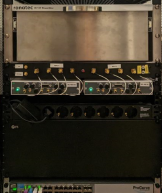}
    \caption{The placement of the devices used for the coordinated location spoofing in NSS lab environment.}
    \label{fig:datasetb}
\end{figure}

\begin{figure}
    \centering
    \includegraphics[width=\columnwidth]{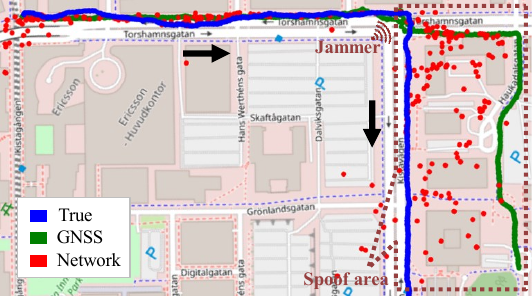}
    \caption{The actual walking trace, \ac{gnss} positions, and network-based positions. \ac{gnss} is spoofed, and network signals (cellular, Wi-Fi, and Bluetooth) are replayed in the red dotted box.}
    \label{fig:traceb}
\end{figure}

\subsection{Dataset B: Coordinated Attack}
The coordinated location spoofing was conducted in our lab environment, mainly utilizing our in-house Skydel \ac{gnss} signal simulator and Ranatec RF shielded box. This setting manipulated wireless signals in the shielded box, while ensuring there is no unauthorized interference with actual users outside the lab facility. 

\subsubsection{Experimental Setup}
\label{subsec:expsetb}
We examined three walking traces, each approximately 1.5 kilometers in length, in Kista Science City and the streets of central Stockholm. The rationale of the test is to record \ac{gnss} and network signals of both benign and attack traces in advance and then superimpose \ac{gnss} spoofing, benign opportunistic ranging data, and adversarial ranging signals (collected via a dedicated trace at a distinct location/path, without any actual adversarial transmissions, thus no adverse effect). The simulator setup includes a workstation with Skydel 24.9.0, a Safran CDM-5 clock distribution module for synchronization, two Ettus USRP N310 \acpl{sdr}, a Ranatec RI 187 RF shielded box, and a Tallysman TW7900P passive triple band \ac{gnss} antenna. Benign, spoofing, and jamming \ac{gnss} signals for \ac{gps} L1, Galileo E1, and BeiDou B1 are generated using Skydel, which streams IQ data to \acpl{sdr}. N310 \ac{sdr} outputs benign, spoofing, and jamming signals with gains of 60/65, 70, and 60 dB, respectively. A 40 dB attenuation is between the connection of the \ac{sdr} output and the antenna input. Then, the antenna emits the generated \ac{gnss} signals in the shielded box. The placement of the devices is shown in Figure~\ref{fig:datasetb}. The jammer enables CW, Chirp, Pulse, BPSK, BOC, and AWGN jamming at the central frequency of 1575.42 MHz with -15 dBm reference power. The spoofer uses -37 dBm reference power. 

As shown in Figure~\ref{fig:traceb}, a jammer is simulated as placed in the middle of the victim’s walking trace, where the actual and the spoofed traces intersect. A spoofer is simulated to closely follow the victim on the actual walking trace. The attack strategy uses the static \ac{gnss} jammer to force the victim receiver to lose lock, and then, jamming is stopped and spoofing is launched for multiple constellations (\ac{gps} L1, Galileo E1, and BeiDou B1). Simultaneously, network-based positioning is targeted in a coordinated manner through the simultaneous, at this point, replay of cellular, Wi-Fi, and Bluetooth signals. These spoofed signals are introduced at the place where the true and spoofed positions initially coincide and then as they gradually deviate. Throughout the attack, benign network signals remain present. The actual trace of the victim in Figure~\ref{fig:traceb} is the blue one, and the attacker imposes the misperception that the victim follows the green path by replaying actually recorded signals while walking across the green path. This can remain unperceived by a user for a significant amount of time. 

\subsubsection{Dataset Collection}
We used a Redmi 9 Android smartphone to collect data. In the pre-collection phase, the GNSSLogger application was used to record \ac{gnss} traces along with onboard sensor measurements, and the NetworkSurvey application recorded messages from cellular, Wi-Fi, and Bluetooth. The sampling rate and format were consistent with Dataset A. In the post-collection phase, attacks were simulated using the smartphone and Skydel within the lab environment. Skydel replays the benign \ac{gnss} signals and generates spoofing signals according to the pre-collected traces. The GNSSLogger application records \ac{gnss} again in the shielded box. For network signals, the pre-collected network messages corresponded to the expected spoofing trace and were timewise aligned to the benign trace. We incorporate network messages of the spoofing trace into the benign trace emulating the attacker replaying network signals. The acquisition of satellite positions and the positions of \acpl{bs}, \acpl{ap}, and Bluetooth anchors was performed in the same way as Dataset A. 

\subsection{Performance of Attack}
During Jammertest 2024, not only were \ac{gnss} receivers affected, but also critical infrastructure, such as cellular \ac{bs} timing, and crowdsourced geolocation databases experienced failures. Network-based positioning dependent on crowdsourcing exhibited significant deviations: Most benign errors result in positions within 200 meters of the ground truth, whereas attack-induced errors predominantly range from 600 to 700 meters. 

As we drove away from the jammer and spoofer, being protected by the terrain/buildings, we observed the affected receivers reacquired positions. Once we moved back to the jammer or spoofer line of sight, the receivers were once again affected. Furthermore, we observed that even when the attack targeted only a single constellation (e.g., \ac{gps}), its signals could impact antenna gain and then disable other constellations. 

During our coordinated location spoofing experiments, with walk-based measurements and in-lab \ac{gnss} simulation, the simulated \ac{gnss} jammer and spoofer were static, with their path loss modeled. When the smartphone moved closer to the jammer, the GNSSLogger application indicated a decreasing carrier-to-noise power ratio. When the smartphone was around 10 meters from the jammer, its \ac{gnss} receiver completely lost lock on the benign signals. Then, we turned off the jammer and the spoofer began transmitting the generated \ac{gnss} signals corresponding to a falsified path. Due to the higher power level of the spoofing signals compared to benign ones, the smartphone \ac{gnss} acquired and locked onto the spoofing signals. The network signal spoofer was emulated at the data level by modifying recorded network traces. Cellular, Wi-Fi, and Bluetooth data from the benign and spoofing traces were merged together in the dataset, causing the network positions from the geolocation algorithm to deviate toward the intended spoofing trace, as per Figure~\ref{fig:traceb}. 

Although the strategy using a \ac{gnss} jammer and \ac{gnss} spoofer demonstrated a high success rate in our experiments, we found it challenging to seamlessly spoof the smartphone without prior jamming but rather gradually increasing the spoofer signal power and drifting the receiver signal tracking. 

\subsection{Performance of Detection}
The attack detection accuracy is evaluated by the true positive rate ($P_\text{tp}$) and the false positive rate ($P_\text{fp}$). $P_\text{tp}$ represents the number of time intervals in which $\mathbf{p}_{\text{lbs}}(t)$ is under attack and correctly detected to be so, over the total number of attack time intervals. Conversely, $P_\text{fp}$ is the percentage of time intervals misclassified as being under attack while this is not so. Additionally, we define $\Delta T_\text{d}$ as the average attack detection latency, capturing the time elapsed from the onset of an attack to its detection. 

Our baseline detection methods include one based on the Google Play Services fused location, the secure fusion-based \ac{gnss} attack detection in \cite{LiuPap:C23}, one based on a Kalman filter, and one based on network-provided position. Google Play location is the most widely used fused location provider on Android devices. Secure fusion \cite{LiuPap:C23} assumes \ac{gnss} is possibly under attack and fuses \ac{gnss}, network, and onboard sensors. If the distance between this fused position and the raw \ac{gps}-provided position exceeds a threshold, the method flags the raw position as a spoofing. The Kalman filter identifies position anomalies by analyzing the residuals, defined as the difference between the smoothed position and the raw \ac{gps} position. Network-based detection is derived from the geolocation algorithm leveraging network-based position only, also using distance discrepancy-based detection. 

The experimental parameters were set as follows: $w=20$, with a regression kernel $K_\text{loc}(t-t')=\exp \left(-0.3(\frac{t-t'}{w})^2\right)$, a second-order polynomial regression model, and a sampling rate set to 50\%. Spoofing labels are determined based on deviations: A positive label is assigned if the distance between the \ac{gnss} position and the ground truth is larger than 30 meters. 

\begin{figure}
    \centering
    \includegraphics[width=.95\columnwidth]{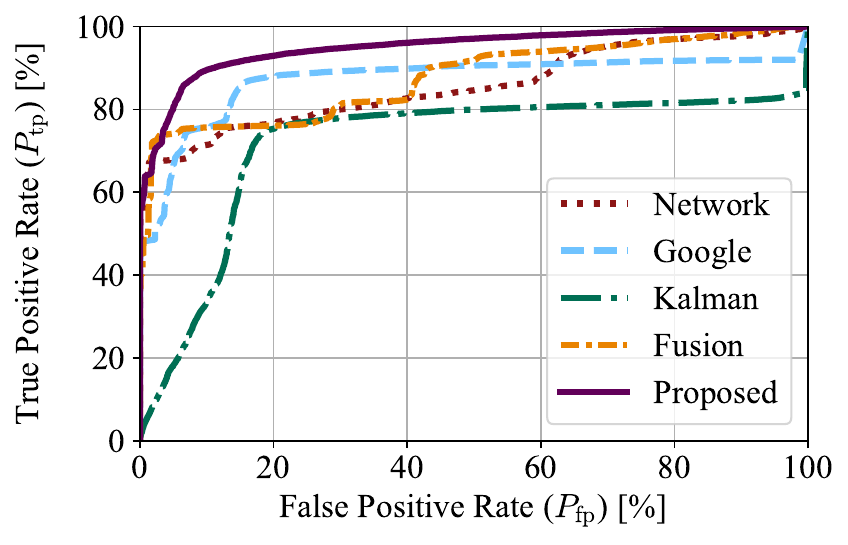}
    \includegraphics[width=.95\columnwidth]{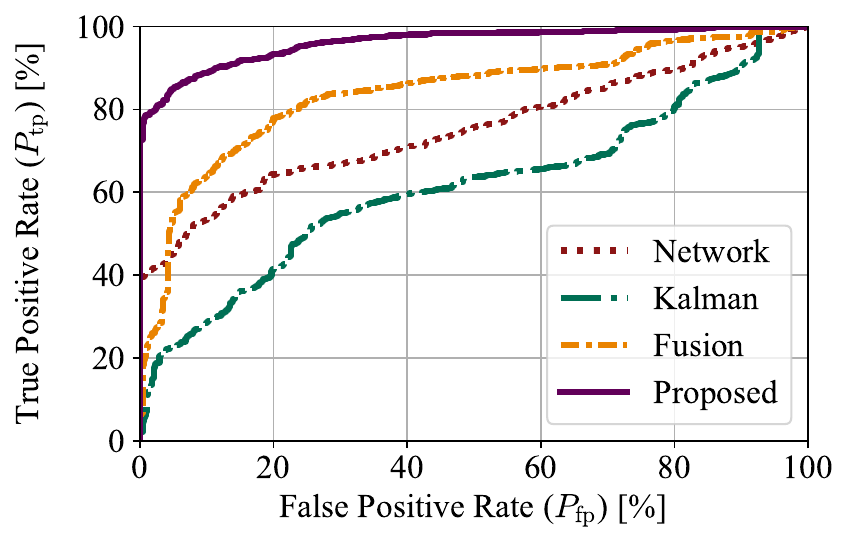}
    \caption{$P_\text{tp}$ of the proposed and baseline methods for Dataset A (upper) and B (lower).}
    \label{fig:resdetexc1}
\end{figure}

Figure~\ref{fig:resdetexc1} presents $P_\text{tp}$ and $P_\text{fp}$ results for Datasets A and B. Our proposed method improves from 18\% to 24\% for Dataset A and up to 58\% for Dataset B in terms of $P_\text{tp}$, compared to Google Play location and network-based detection, when $P_\text{fp}$ is between 5\% and 10\%. This confirms that as long as \ac{gnss} \ac{dop} is such that it leads to positions significantly more accurate than those based on network position, the Google Play location prioritizes \ac{gnss} position even when it is spoofed. It is important to note that Dataset B was collected in the shielded box, so it did not provide valid Google Play locations. Furthermore, network-based positioning lacks precision due to the corruption of geolocation databases caused by spoofing attacks in Dataset A. Additionally, coordinated location spoofing in Dataset B gradually increases the spoofing induced deviation, which limits the network position and Kalman filter based approaches to detect anomalies. Notably, $P_\text{tp}$ is evaluated at the level of individual position fixes rather than over entire traces. $P_\text{tp}$ represents the proportion of correctly identified spoofed intervals among all spoofed intervals, as opposed to $\Delta T_\text{d}$ measuring detection latency per trace.

\begin{figure}
    \centering
    \includegraphics[width=.95\columnwidth]{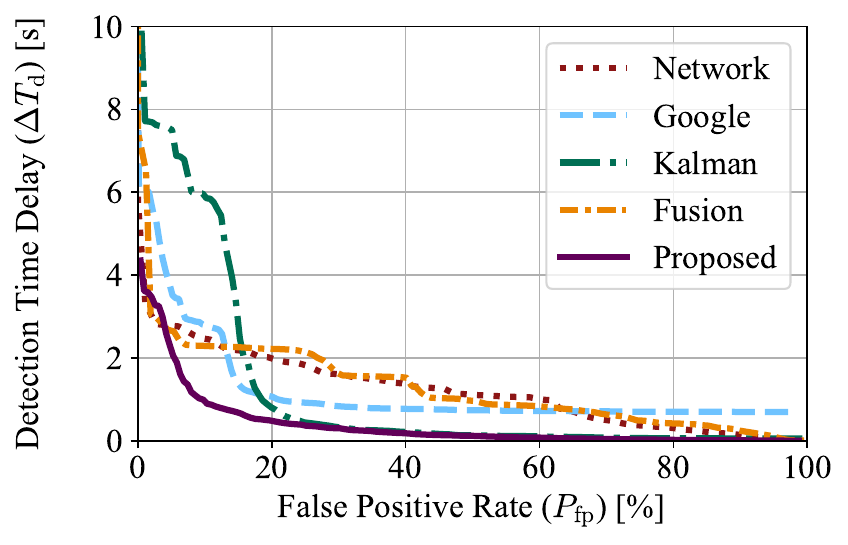}
    \includegraphics[width=.95\columnwidth]{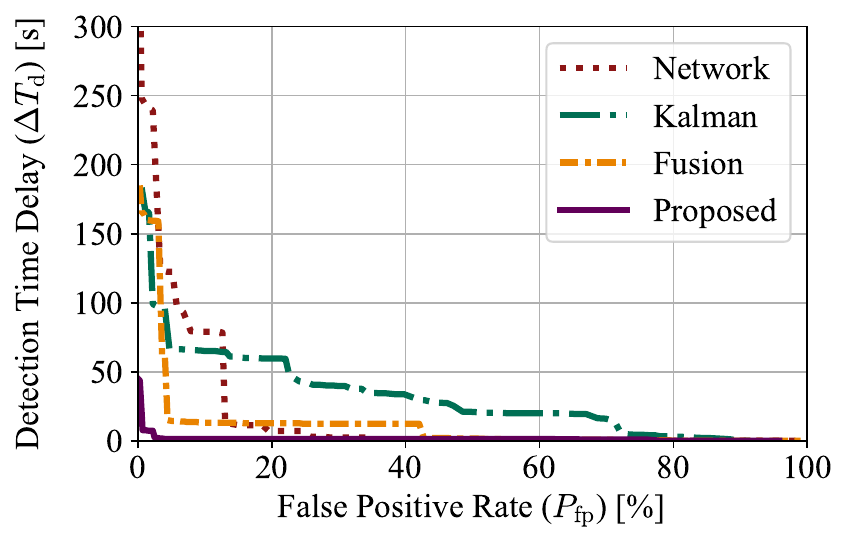}
    \caption{$\Delta T_\text{d}$ of the proposed and baseline methods for Dataset A (upper) and B (lower).}
    \label{fig:resdetexc2}
\end{figure}

Figure~\ref{fig:resdetexc2} shows $\Delta T_\text{d}$ versus $P_\text{fp}$. Our proposed method has up to 5 seconds gain for Dataset A in $\Delta T_\text{d}$, when $P_\text{fp}$ is between 5\% and 10\%. For our proposed detection, $\Delta T_\text{d}$ can detect spoofing attacks within an average of 4 seconds. 

Compared with existing position fusion schemes and products, the proposed scheme can securely fuse position information, detecting and excluding malicious data. Unlike traditional fusion for secure localization \cite{LiuPap:C23}, our scheme adopts a more fine-grained approach by analyzing individual ranging information for enhanced position security. However, the generation of subsets and the computation of positions for cross-validation are time-consuming. Hence, further effort on subset sampling strategies is needed.

\subsection{Scheme Tuning}
We analyze the sensitivity of our scheme to parameters, including sampling rate and window size. For the subset sampling strategy (Section~\ref{subsec:subran}), we evaluate the effect of different sampling ratios on $P_\text{tp}$. The results for Dataset A show that $P_\text{tp}$ increases with higher sampling rates: With an increase from 0.25 to 1.0, $P_\text{tp}$ improves from 86\% to 90\%, at $P_\text{fp}=10\%$. This suggests that higher sampling rates capture a wider range of subsets, while at lower sampling rates the algorithm may miss important subsets. However, the accuracy is relatively insensitive to the sampling rate. When the sampling rate decreases from 1 to 0.25 (4 times less computation), $P_\text{tp}$ is reduced by at most 4\%. Although higher sampling rates generally lead to improved detection accuracy, practical constraints due to client computational resources and acceptable $\Delta T_\text{d}$ are important.

By adjusting $w$ in \eqref{eq:proall}, we evaluate the detection performance for different window sizes. Selecting an appropriate size requires a trade-off, as smaller $w$ requires less computing power but potentially degrades detection accuracy. Conversely, oversized $w$ may result in processing unnecessary historical data and increased computational complexity. Our results indicate relatively stable $P_\text{tp}$ trending for $5<w<25$, but the computational time for detection when $w=25$ is 1.7 times that of $w=5$. In addition, $P_\text{tp}$ for $15<w<25$ exhibits slightly improved performance compared to other window sizes; consequently, we selected this as the preferred range in our detector. We further conduct a small scale comparison among different orders of local polynomial regression in $\mathbf{W}$, showing that $P_\text{tp}$ is higher for $n=2$ compared to $n=1$. 

\section{Discussion}
\textbf{Ethical Concerns and Limitations:} As \ac{gnss} spoofing is illegal, the experiments that led to Dataset A were conducted at Jammertest by the Norwegian authorities. For Dataset B, all spoofing signals were entirely contained within our RF shielded box, ensuring that they had no effect outside the lab. The adversarial trace collection for opportunistic signals was essentially a benign collection along an actual yet labelled and used, or the sake of emulating the attack, adversarial (to be imposed on the victim) path, with a superposition of this data with the benign one. 

\textbf{Alternative Data Cleaning:} We explored \ac{llm}-based data cleaning, since the Wi-Fi \ac{ap} position data is crowdsourced from WiGLE.net \cite{BobArkUht:J23} without quality assurance. The process of matching Wi-Fi \ac{ap} \ac{ssid} to their corresponding place names involves leveraging an \ac{llm} to extract text semantic information and utilizing \ac{poi} APIs for mapping and validating. The following steps outline this data cleaning: Step 1 filters fixed places. It inputs the \ac{ssid} name of the \ac{ap} using the prompt ``Is this Wi-Fi \ac{ssid} from a static or mobile hotspot: \texttt{SSID\_NAME}? Please answer static or mobile only.'' This prompt directs the \ac{llm} to distinguish between fixed and mobile \ac{ap} based on semantics. By filtering out mobile \acpl{ap}, we focus on identifying fixed places. Step 2 extracts keywords. It inputs the \ac{ssid} name of the \ac{ap} using the prompt ``Then, can you extract some keywords of the place name from \texttt{SSID\_NAME}? Please answer keywords directly.'' This step prompts the \ac{llm} to extract relevant keywords indicative of the place name associated with the \ac{ap}. Step 3 queries \ac{poi} API based on the previously extracted keywords are used to obtain \ac{poi} coordinates. 

\textbf{Processing Requirements and Overhead:} We tested different brands of smartphones with chipsets from Exynos, MTK, Qualcomm, and Google Tensor. They all support logging \ac{gnss} pseudoranges at 1 Hz and network survey data every 3--10 seconds, depending on connectivity. Hence, detection should process the \ac{gnss} data within 1 second and the network data within 3 seconds. We ran the Python version of the algorithm on the aforementioned MTK and Google Tensor platforms. Although the computation for the subsets took the longest time, each detection can be completed within 1 second without parallel optimizations. Even if the computing power of the mobile platform were not sufficient, we could still change the sampling rate in Section~\ref{subsec:subran}. 

\textbf{Deployment and Future Work:} Although our experimental results and simulations show practical performance using opportunistic signals and consumer-grade sensors, the attack and detection test scenarios should be extended to cover, e.g., subtler, stealthier attacks, \ac{gnss} cold-start settings, static versus dynamic victim receivers, and scenarios with limited opportunistic information. The attacker can subtly change position, time, or signal power, in spite of the complexity of such attacks. Hence, integrating our detection with other lower-layer (position-, time-, and signal-based) spoofing detectors could provide a multi-layer defense. An attack during a cold-start period could impact \ac{gnss} performance, but our position consistency-based detection can remain robust. Future work should consider various deployment settings, including different hardware, user mobility, topology, and attack strategies.

\section{Related Work}
\label{sec:relwor}
\textbf{\ac{raim} Protecting \ac{gnss}:} There are two primary forms of \ac{raim}: residual-based and solution-separation \cite{JoeChaPer:J14}. Residual-based \ac{raim} uses statistical hypothesis checking at the residual errors to identify potentially inaccurate measurements \cite{KhaRosLanCha:C14,RoyFar:C17}. The residuals can come from the least squares or Kalman filters: \ac{ekf} \ac{raim} makes use of sliding window filters to identify and eliminate outliers using \ac{gps} and inertial sensors \cite{KhaRosLanCha:C14,RoyFar:C17}. Solution-separation \ac{raim} recursively assumes faulty satellites, generates subsets of the remaining satellites to derive solutions, and then excludes faults \cite{ZhaPap:C19,LiuPap:C24}. For example, \cite{ZhaPap:C19} integrates RANSAC clustering to classify position solutions. 

\textbf{Rogue Wi-Fi \ac{ap} Detection:} Detection of malicious Wi-Fi hotspots has received increasing interest. Some popular industry solutions \cite{Aru:J24,Ibm:J24,Gan:J24} ignore all unknown \acpl{ap} or use \ac{ap} \ac{mac} and \ac{ssid} whitelists to prevent unauthorized \acpl{ap}. However, attackers can forge these records rather effortlessly. For example, most of the consumer-grade Wi-Fi routers can set any \ac{ssid}, and open-source routers (e.g., OpenWrt) can even modify their \ac{mac} addresses. Hence, there is a need for detection beyond these Wi-Fi beacons. \cite{LinGaoLiDon:C20} uses wireless fingerprinting technology independent of client devices, but the robustness of fingerprinting gets worse in dynamic environments (rain, high traffic, etc.). In addition, semantic-based \ac{csi} in \ac{iot} environments offers potential accuracy advantages but requires specialized hardware and large-scale frequency band scanning \cite{BagRoeMarSch:C15}. 

\section{Conclusion}
\label{sec:conclu}
Our extended \ac{raim} scheme detects \ac{lbs} position manipulation using opportunistic ranging information and onboard sensor measurements: It cross-validates position estimates from different sources to give an attack likelihood. We implement position attacks in real-world \ac{lbs} applications and show the feasibility of the proposed detection with experiments in various scenarios. The benefit is a significant reduction of both user and service exposure to scams without adding additional hardware.

\begin{acks}
This work was supported in part by the SSF SURPRISE cybersecurity project, the Security Link strategic research center, and the China Scholarship Council. The computational resources were provided by NAISS, partially funded by the Swedish Research Council through grant agreement no. 2022-06725. We would also like to acknowledge the work of the organizers of Jammertest 2024. 
\end{acks}

\bibliographystyle{ACM-Reference-Format}
\balance
\bibliography{reference/references}

\appendix
\section{Wi-Fi Spoofing with GNSS Jamming}
\label{app:appenda}
\begin{figure}
    \centering
    \includegraphics[trim={0 30cm 0 30cm},clip,width=.31\columnwidth]{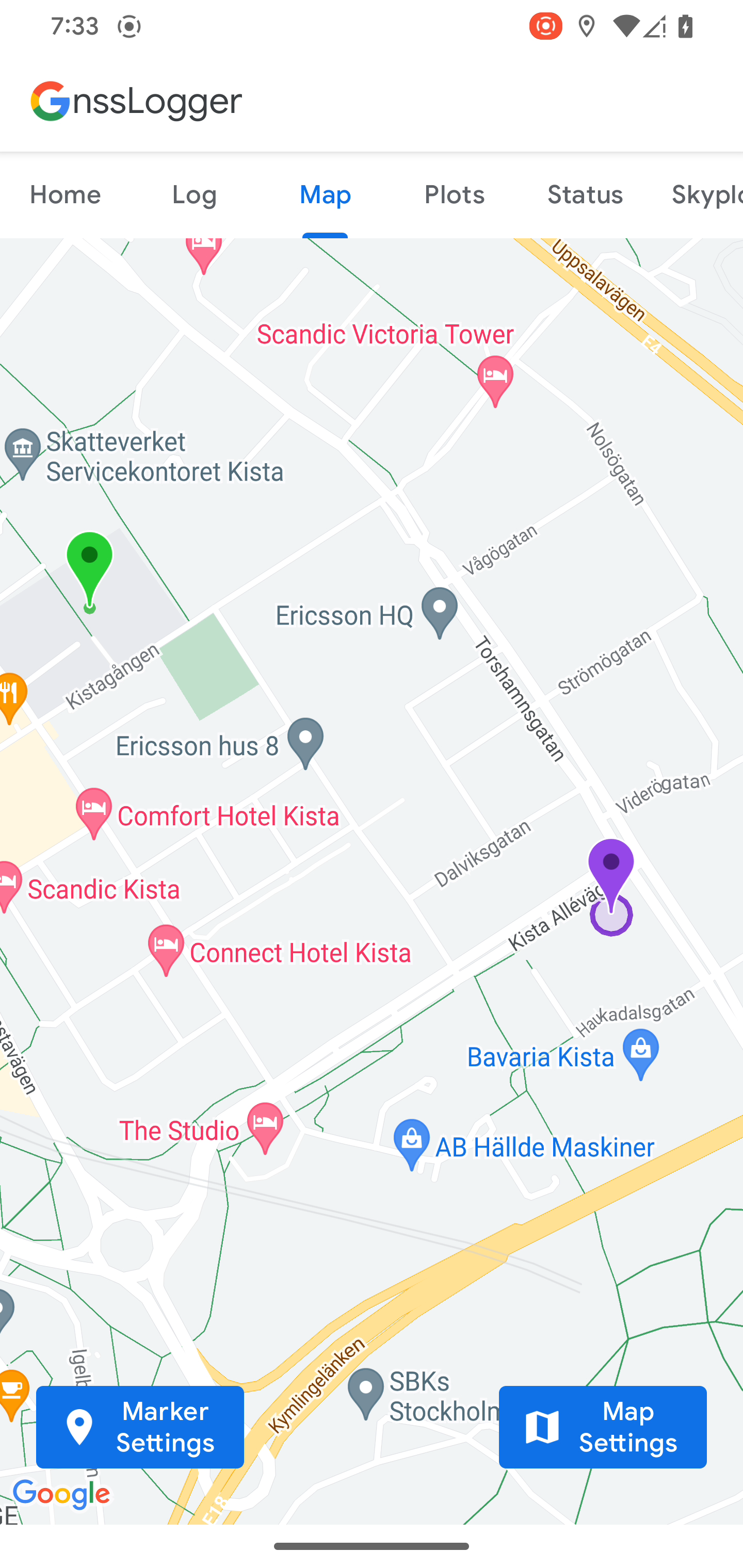}
    \includegraphics[trim={0 30cm 0 30cm},clip,width=.31\columnwidth]{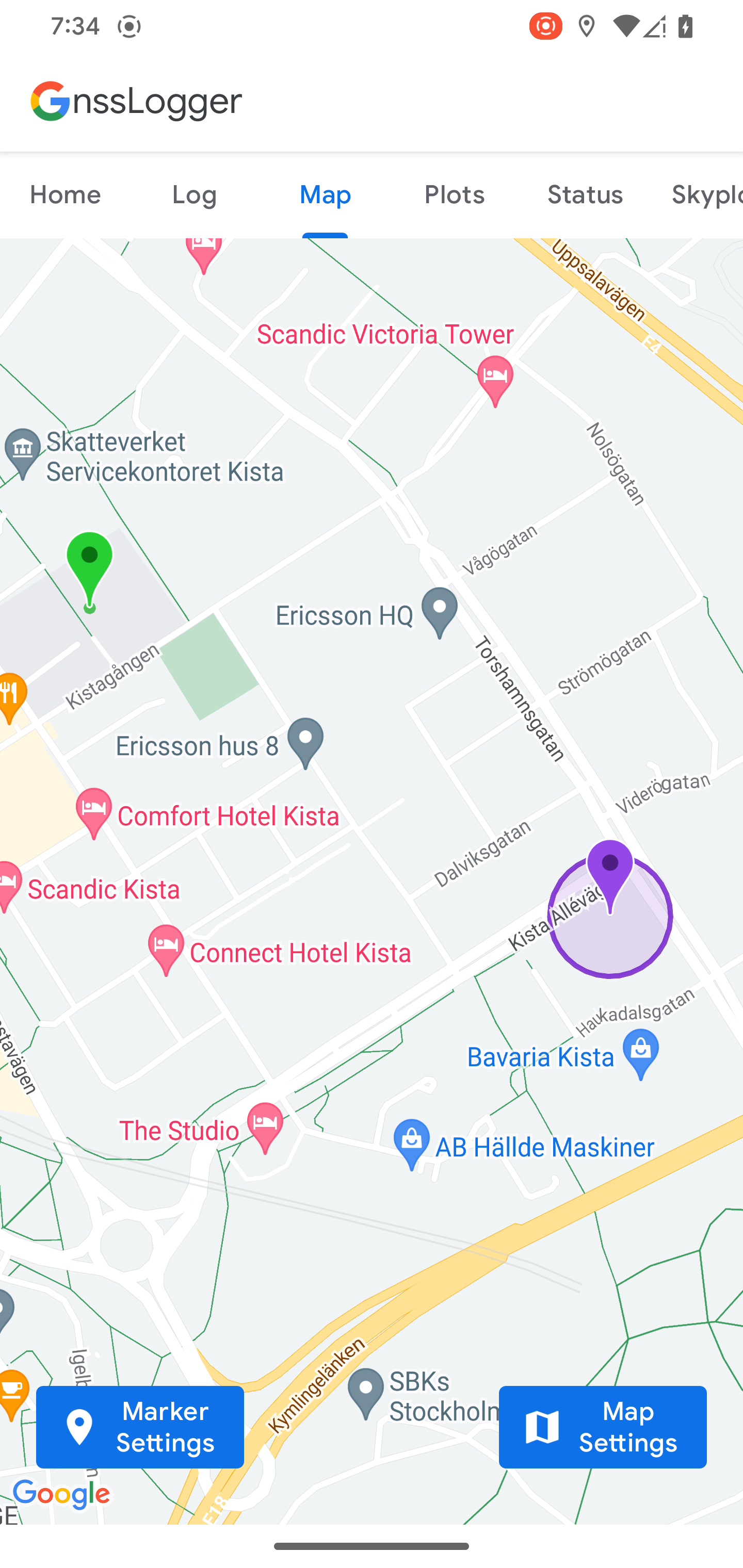}
    \includegraphics[trim={0 30cm 0 30cm},clip,width=.31\columnwidth]{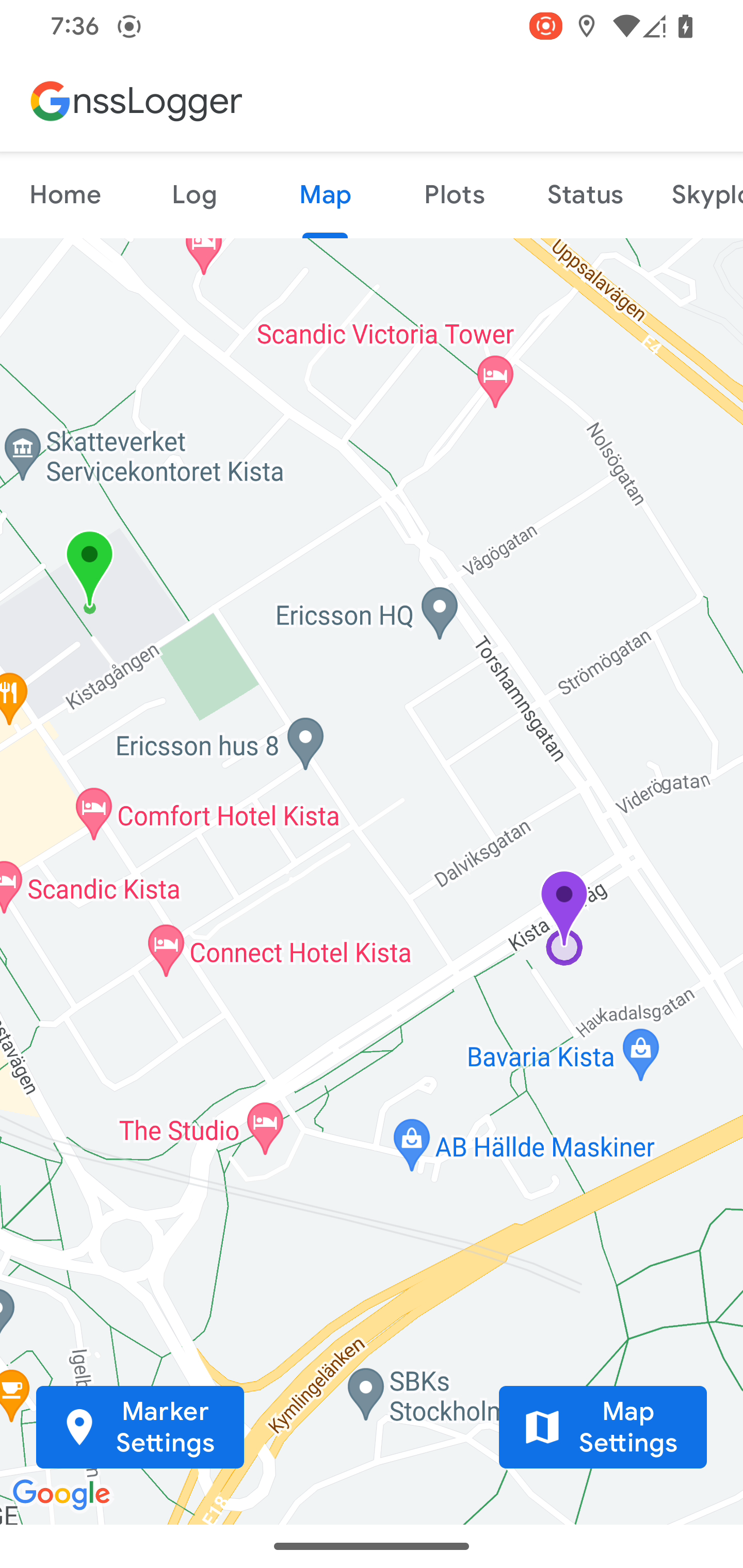}
    \includegraphics[trim={0 30cm 0 30cm},clip,width=.31\columnwidth]{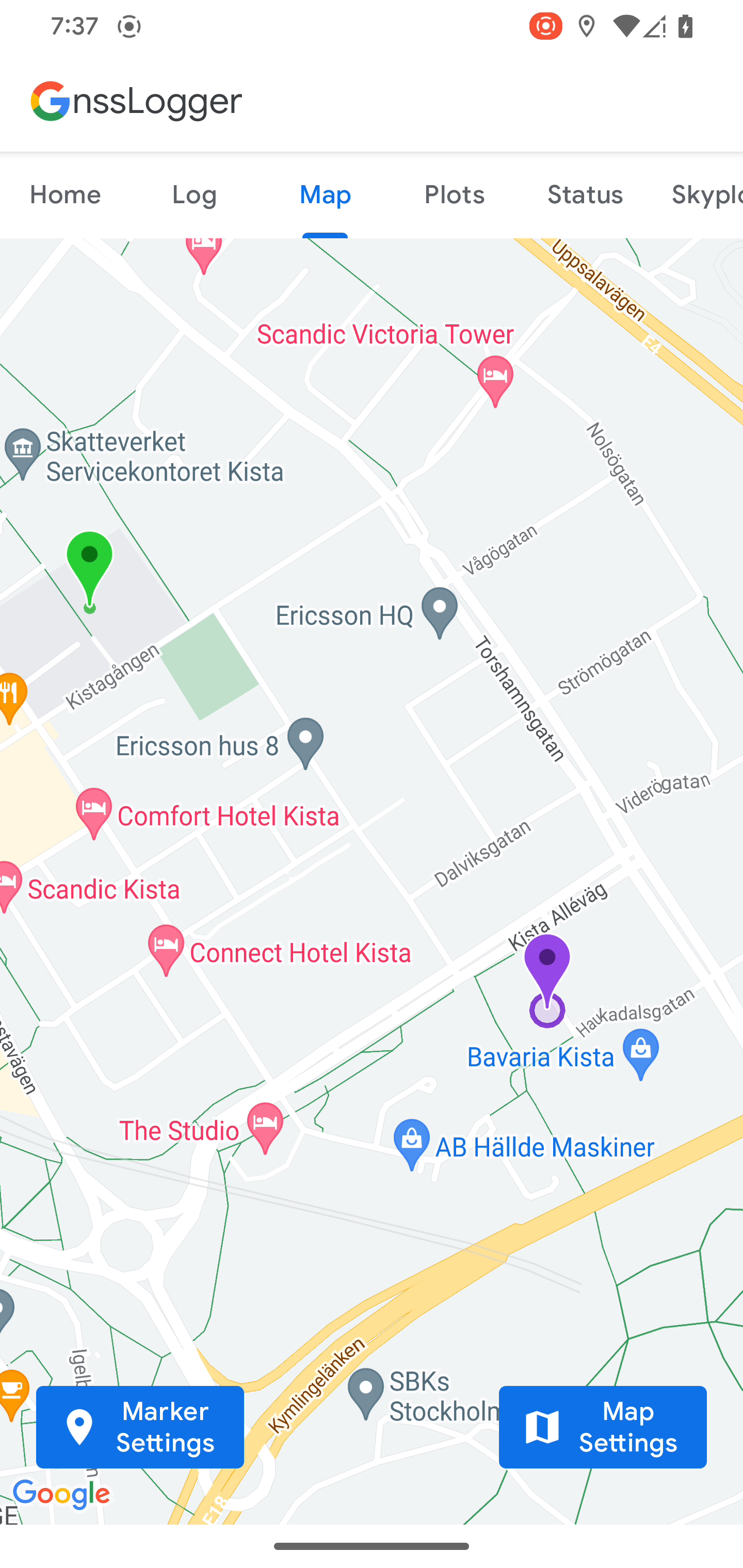}
    \includegraphics[trim={0 30cm 0 30cm},clip,width=.31\columnwidth]{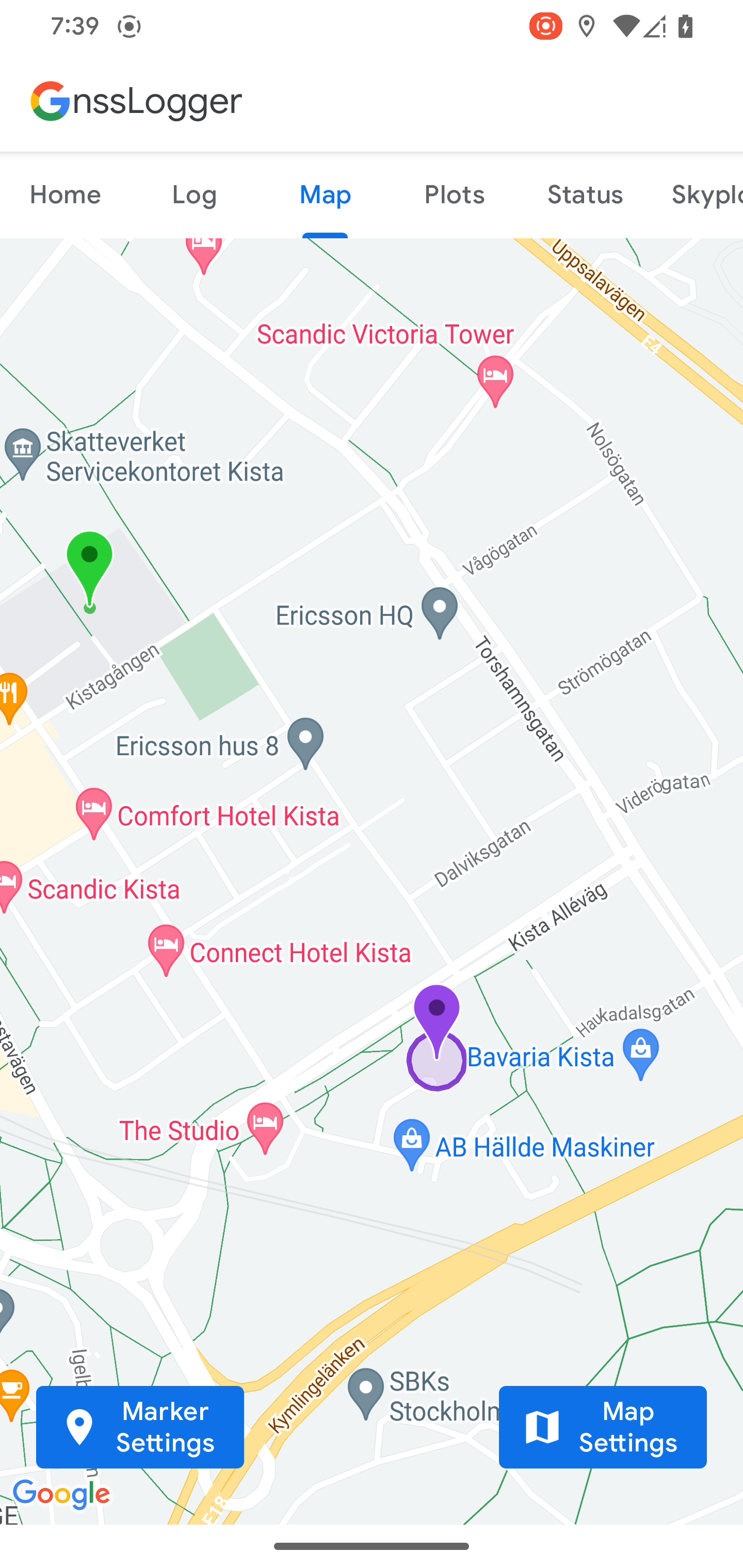}
    \includegraphics[trim={0 30cm 0 30cm},clip,width=.31\columnwidth]{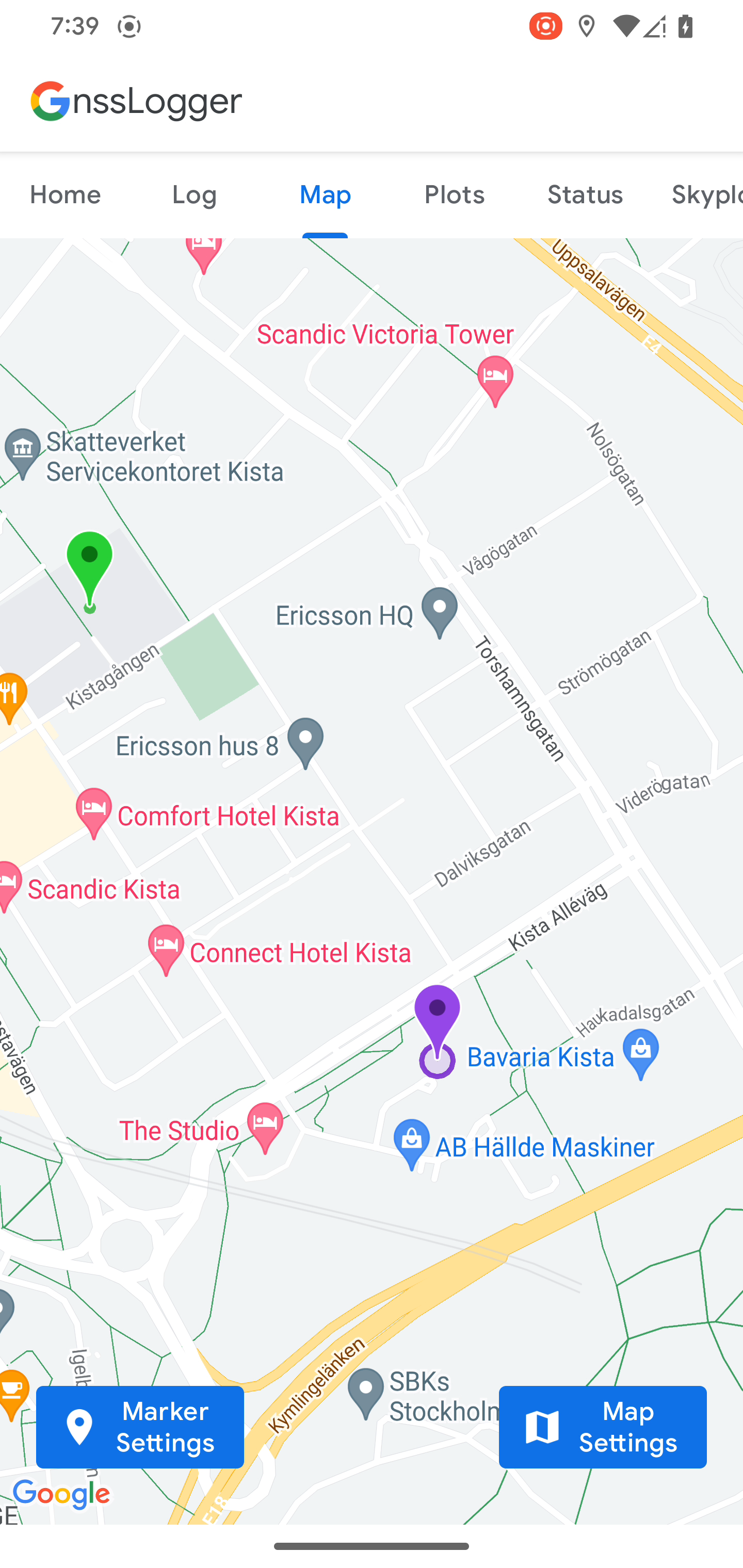}
    \caption{A time series of screenshots (from left to right, top to bottom) of a recorded attack video.}
    \label{fig:appdegatt}
\end{figure}

Figure~\ref{fig:appdegatt} shows an application using the map service is fooled by a Wi-Fi router broadcasting beacons based on a list of \ac{ssid} and \ac{bssid} along a pre-selected road. The purple pin is the estimated position from the application, but actually, the smartphone is located at the position of the dark green pin. We have also made two screen recordings demonstrating position manipulation under real \ac{gnss} jamming and Wi-Fi replaying in Bleik (available at: \url{https://drive.google.com/drive/folders/1rZtwVYXi3OwKyS8Yzn23E2E18rBP-J3x}).

\begin{lstlisting}[language=python, basicstyle=\footnotesize, breaklines=true, caption=Wi-Fi location spoofing script tested on a consumer-grade router installed OpenWrt system and enabled virtual \acpl{ap}. It will broadcast Wi-Fi beacons of \texttt{ap\_set}.]
def uci_set_wifi(ap_set):
    ssh.exec_command("sed -i '37,$d' /etc/config/wireless")
    for i in range(min(len(ap_set), 15)):
        cmd_to_execute = "uci batch << EOF\n"
        cmd_to_execute += f"""
            set wireless.wifinet{i}=wifi-iface
            set wireless.wifinet{i}.device='radio1'
            set wireless.wifinet{i}.mode='ap'
            set wireless.wifinet{i}.ssid='{ap_set.loc[i, "SSID"]}'
            set wireless.wifinet{i}.encryption='psk2'
            set wireless.wifinet{i}.macaddr='{ap_set.loc[i, "BSSID"]}'
            set wireless.wifinet{i}.key='Password'
            """
        cmd_to_execute += "commit\nEOF"
        ssh.exec_command(cmd_to_execute)
    ssh.exec_command("wifi reload")
\end{lstlisting}

\section{Coordinated Location Spoofing}
\label{app:appendb}
\begin{figure}
    \centering
    \begin{subfigure}{\columnwidth}
        \includegraphics[trim={0 3cm 0 3.1cm},clip,width=\columnwidth]{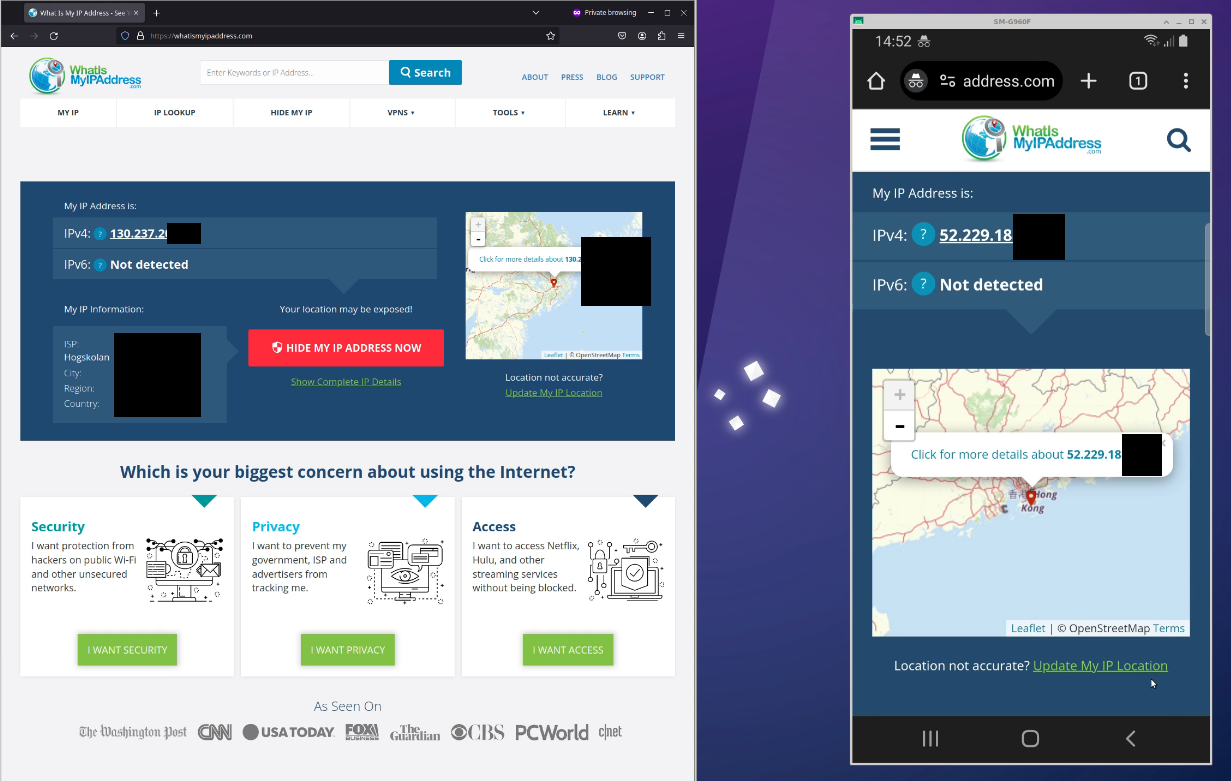}
        \caption{GeoIP manipulation by a Wi-Fi router relaying all TCP and UDP messages. Left: actual IP; right: manipulated.}
    \end{subfigure}
    \\
    \begin{subfigure}{.6\columnwidth}
        \includegraphics[width=\columnwidth]{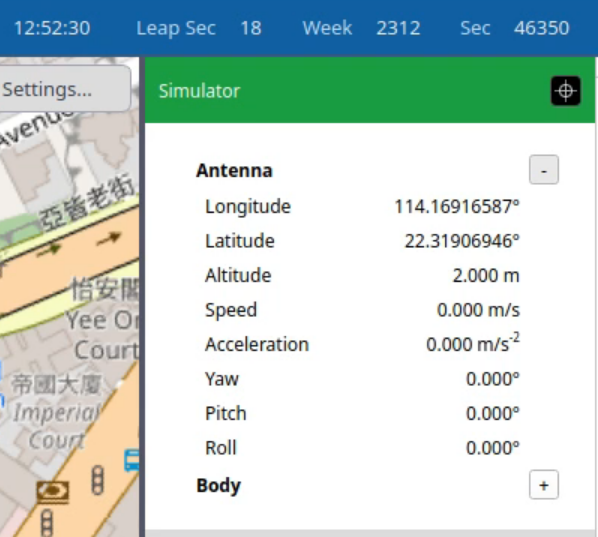}
        \caption{\Ac{gnss} signal generator setting.}
    \end{subfigure}
    \hfill
    \begin{subfigure}{.35\columnwidth}
        \includegraphics[trim={0 0 0 3.2cm},clip,width=\columnwidth]{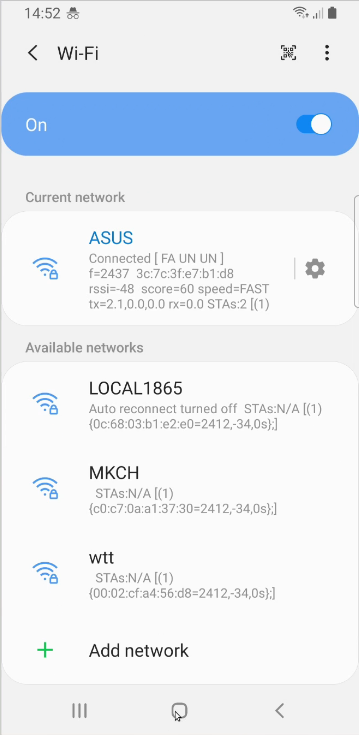}
        \caption{Generated beacons.}
    \end{subfigure}
    \caption{Settings of a coordinated attack.}
    \label{fig:appcoratt}
\end{figure}

This attack includes packets relaying for GeoIP manipulation, Wi-Fi location spoofing, and Skydel-based \ac{gnss} spoofing, as in Figure~\ref{fig:appcoratt}. The spoofed positions are coordinated, meaning the position of the relaying server and the coordinates in \ac{gnss} signal generator setting are near the pre-selected spoofed position. Moreover, the generated beacons mimic \ac{ssid} and \ac{bssid} of Wi-Fi beacons. 

\begin{lstlisting}[language=bash, basicstyle=\footnotesize, breaklines=true, caption=The iptables rules for a proxy server running on the IP \texttt{127.0.0.1} and port \texttt{12345}.]
ip rule add fwmark 1 table 100 
ip route add local 0.0.0.0/0 dev lo table 100
iptables -t mangle -N LAN
iptables -t mangle -A LAN -d 127.0.0.1/32 -j RETURN
iptables -t mangle -A LAN -d 224.0.0.0/4 -j RETURN 
iptables -t mangle -A LAN -d 255.255.255.255/32 -j RETURN 
iptables -t mangle -A LAN -d 192.168.0.0/16 -p tcp -j RETURN
iptables -t mangle -A LAN -d 192.168.0.0/16 -p udp ! --dport 53 -j RETURN
iptables -t mangle -A LAN -j RETURN -m mark --mark 0xff
iptables -t mangle -A LAN -p udp -j TPROXY --on-ip 127.0.0.1 --on-port 12345 --tproxy-mark 1
iptables -t mangle -A LAN -p tcp -j TPROXY --on-ip 127.0.0.1 --on-port 12345 --tproxy-mark 1
iptables -t mangle -A PREROUTING -j LAN
\end{lstlisting}

\end{document}